 \documentclass{aa}  
\usepackage{graphicx}
\usepackage{times}
\usepackage{xspace}
\usepackage{epsfig}
\usepackage{natbib}
\usepackage{rotating}
\usepackage{dcolumn}
\usepackage{txfonts}
\usepackage{graphicx}
\usepackage{lscape}
\usepackage{txfonts}
\begin{document}
   \title{The seven sisters DANCe}
   \subtitle{II: Proper motions and the Lithium-Rotation-Activity connection for G and K Pleiads}

\author{D. Barrado\inst{1}
          \and
          H. Bouy
          \inst{1}
          \and
	  J. Bouvier
          \inst{2}
          \and
          E. Moraux
          \inst{2}
          \and
          L. M. Sarro
          \inst{3}
          \and
          E. Bertin
          \inst{4}
          \and 
          J.C. Cuillandre
          \inst{5}
          \and
          J.R. Stauffer
          \inst{6}
          \and
          J. Lillo-Box
          \inst{7}
          \and
          A. Pollock
          \inst{8}
}

   \offprints{D. Barrado}

\institute{
  Depto. Astrof\'{\i}sica, Centro de Astrobiolog\'{\i}a (INTA-CSIC),  ESAC campus,
  Camino Bajo del Castillo s/n,
  E-28692 Villanueva de la Ca\~nada, Spain %\\  \email{D. Barrado, barrado@cab.inta-csic.es}
         \and
             UJF-Grenoble 1/CNRS-INSU, Institut de Plan\'etologie et d’Astrophysique de Grenoble (IPAG), UMR 5274, Grenoble, F-38041, France%\\
         \and
             Dpto. de Inteligencia Artificial, ETSI Inform\'atica, UNED, Juan del Rosal, 16, E-28040, Madrid, Spain
         \and
             CEA/IRFU/SAp, Laboratoire AIM Paris-Saclay, CNRS/INSU, Université Paris Diderot, Observatoire de Paris, PSL Research University, F-91191 Gif-sur-Yvette Cedex, France
         \and
             CEA/IRFU/SAp, Laboratoire AIM Paris-Saclay, CNRS/INSU, Université Paris Diderot, Observatoire de Paris, PSL Research University, F-91191 Gif-sur-Yvette Cedex, France 
         \and
             Spitzer Science Center, California Institute of Technology, Pasadena, CA 91125, USA.%\\
         \and
             European Southern Observatory, Alonso de Cordova 3107, Vitacura Casilla 19001, Santiago 19, Chile
         \and
             European Space Agency XMM-Newton Science Operations Centre, European Space Astronomy Centre, P.O. Box 50727, Villafranca del Castillo, E-28080 Madrid, Spain
}

\titlerunning{Lithium, rotation and activity in G and K Pleiads}

   \date{Received , 2014; accepted }

\abstract
% context heading (optional)
{
Stellar clusters are open windows to understand stellar evolution. Specifically, the change with time and the dependence on mass of different stellar properties.
As such, they are our laboratories where different theories can be tested.}
% aims heading (mandatory)
{
We try to understand the origin of the connection between lithium depletion in F, G and K stars, rotation and activity, in particular in the Pleiades 
open cluster.
}
% methods heading (mandatory)
{
We have collected all the relevant data in the literature, including information regarding rotation period, binarity and activity,  and cross-matched with  proper motions,  multi-wavelength photometry and membership probability from  the DANCe database. In order to avoid biases, only Pleiades single members with probabilities larger than 75\% have been included in the discussion.
}
% results heading (mandatory)
{
The analysis confirms that there is a strong link between  activity, rotation  and the lithium equivalent width excess, specially for the 
range  $Lum(bol)=0.5-0.2$ $L_\odot$ (about K2-K7 spectral types  or 0.75--0.95 $M_\sun$). 
}
% conclusions heading (optional)
{
 It is not possible to disentangle these effects but we cannot exclude that the observed lithium overabundance is partially an observational effect due to enhanced activity, due to a large coverage  by stellar spots induced by high rotation rates.
Since a {\it bona fide} lithium enhancement is present in young, fast rotators, both activity and rotation should play a role in the lithium problem.}


\maketitle

%%%%%%%%%%%%%%%%%%%%%%%%%%%%%%%%%%%%%%%%%%%%%%%%%%%%%%%%%%%%%%
%%%%%%%%%%%%%%%%%%%%%%%%%%%%%%%%%%%%%%%%%%%%%%%%%%%%%%%%%%%%%%
%%%%%%%%%%%%%%%%%%%%%%%%%%%%%%%%%%%%%%%%%%%%%%%%%%%%%%%%%%%%%%
%%%%%%%%%%%%%%%%%%%%%%%%%%%%%%%%%%%%%%%%%%%%%%%%%%%%%%%%%%%%%%
%%%%%%%%%%%%%%%%%%%%%%%%%%%%%%%%%%%%%%%%%%%%%%%%%%%%%%%%%%%%%%
\section{Introduction}\label{sect:intro}

The Pleiades open cluster does not only offer us a  beautiful spectacle during the Fall, 
it is one of the best studied stellar associations
and one of the cornerstones in order to understand stellar properties and  evolution.
In fact, the literature includes more than one thousand refereed papers dealing with the Pleiades, 
most of them using the Pleiades as a reference, only in the last ten years.
In spite of this, the Pleiades cluster still keeps many secret and basic parameters, 
such as  its distance and age, are not unambiguously established.
Even its census is incomplete, although the recent works by \cite{Stauffer2007.1},
\cite{Lodieu2012.Pleiades}
 and  
\cite{Bouy2013-DANCE}  
 have improved considerably the membership list,
In fact,   
\cite{Bouy2015-Pleiades}  
has increased  the number
 of known members by a 50\%,  by using public archival data,
  very accurate proper motions and multi-wavelength photometry 
(see additional details in \citealp{Sarro2014.1}). 
Regarding its distance, there are currently two different methodologies, based,
 respectively, on parallaxes from Hipparcos (\citealp{Perryman97})
 and isochrone fitting.
 Pre-Hipparcos distances  for the Pleiades range between 125 and 130 pc (see, for instance, \citealp{Soderblom1993-RotationActivity-FGK-Pleiades}),  whereas the initial distance derived by Hipparcos is much closer, about 119 pc (\citealp{vanLeeuwen1999-Hipparcos-9clusters}). This last value is significantly different to the distance  derived by \cite{Pinsonneault1998.1}, who used  color-magnitude diagrams and fitting isochrones  and obtained $133.5\pm1.2$  pc.
 More recently,
 \cite{vanLeeuwen2009.1}, by re-analyzing Hipparcos data, has derived a 
distance of  $120.2\pm1.9$ pc.
Note that these values should correspond to the distance to the cluster center,
 whose  core radius should be  around 3 degrees, which corresponds to 5-6 pc.
 Another trigonometric parallax, based on HST data and three members,
 comes from \cite{Soderblom2005-PleiadesDistance},
which produced  $134.6\pm3.1$ pc.
 Lately,  \cite{Melis2014.1} have derived $136.2\pm1.2$ pc based on an accurate  parallax for four bona-fide  members
obtained with the VLBI, which also agrees with the value derived by Galli et al. (2016, in prep), by using
accurate proper motions and the convergence point method ($137.7\pm2.5$ pc).

On the other hand, the evolution of lithium, either from the cosmological or from the 
stellar perspective, has received a significant amount of attention since it gives us 
access to the early universe, the late evolution of stars or their internal structure.
 In this last case, since its abundance depends on stellar age (\citealp{Herbig1965-Li-FG}), it has been used 
as an evolutionary tracker. However, we are far from understanding all sides of the lithium problem.

For FGK members belonging to  open clusters there is a clear dependence of the lithium abundance 
 with mass, age and other parameters such as rotation/activity. 
In fact, standard models predict that the depletion happens during the Pre-Main Sequence
evolution. However, the observations show that the depletion continues beyond the arrival to the ZAMS, 
so additional, non-standard mixing has to take place.
Moreover,  for clusters older than the Pleiades there is a narrow effective temperature
range (6400-6900 K) which shows a large depletion of lithium abundance due to
non-standard mixing, the so 
called lithium gap, dip or chasm  (\citealp{Boesgaard1986-Li-Fdwarfs}; \citealp{Michaud1991-Li}; 
\citealp{Balachandran1995-Li-F-M67}).
In any case, the complexity of the evolution has been established by multiple studies focusing on
clusters of different ages. Seminal papers, to name a few,  are
\cite{Pilachowski1984-Li-NGC7789},  \cite{Pilachowski1986-Li-NGC7789},
\cite{Pilachowski1987-Li-F-Pleiades},  \cite{Pilachowski1988-L-NGC752-M67}, 
\cite{Pasquini1997-Lithium-M67}, \cite{Pasquini2008-M67-Li}
for NGC7789, the Pleiades, NGC752, and M67;
\cite{Boesgaard1986-Li-Hyades},  \cite{Boesgaard1987-Li-Coma}, \cite{Boesgaard1987-Li-F-Hyades}, 
 \cite{Boesgaard1988-Li-Pleiade-APer},  \cite{Boesgaard1988-Li-HyadesMG-Praesepe},  
\cite{Boesgaard1989-LiBe-Hyades}, \cite{Thorburn1993-Li-Hyades},  \cite{Barrado1996-Li-Hyades})
  for the Hyades, Coma,  the Pleiades and Alpha Per,  and Praesepe;
\cite{Soderblom1993-Li-Praesepe}, \cite{Soderblom1993-Li-Pleiades}, \cite{Soderblom1993-Li-UMaG} 
for Praesepe, the Pleiades and Ursa Majoris moving group.
More recently, additional observations for  clusters, generally younger,
 have been studied. Again, just to provide some references: 
NGC2516 and M35, almost Pleiades twins 
(\citealp{Jeffries1998-LiRotation-NGC2516}; \citealp{Barrado2001-Li-M35}),
IC2602 and IC2391 (\citealp{Barrado1999-Li-LDB-IC2391}; \citealp{Randich2001-Li-K-IC2602}; 
\citealp{Randich2001_Li_Metal_IC2602_IC2391}; \citealp{Barrado2004-Li-Ha-IC2391}),
NGC2547 (\citealp{Jeffries2003-Li-NGC2547}),
IC4665 (\citealp{Jeffries2009-Li-IC4665}), and
Collinder 69 (\citealp{Dolan1999-Li-C69}, \citealp{Bayo2012-Li-Rotation-Activity-C69}).
In the very near future, the large scale spectroscopic survey Gaia-ESO 
(\citealp{Gilmore2012.GaiaESO}; \citealp{Randich2013.GaiaESO})
 will provide an extended database. Recent example is the case of the Vela OB2 association 
 (\citealp{Sacco2015.VelaOB2}).

Lately, several works have been published trying to understand the lithium content in solar-type stars from different perspectives.
\cite{Bouvier2008.1} --see also \citealp{Eggenberger2012-Li-Rotation-Disk-PMS}-- 
investigates the effect of the disk life-time on the rotation and lithium:
 slow rotation would be the consequence of long-lasting star-disk interaction during the PMS and would produce a significant decoupling 
between the core and the convective envelope, with the final consequence of extra-mixing and higher lithium depletion.
The analysis by \cite{Bouvier2016-Li-Rotation-NGC2264} of the Gaia-ESO data corresponding to
5 Myr old cluster  NGC226   shows a lithium enhancement for
 fast rotators in with effective temperature in the range 3800-4400 K.
On the other hand, \cite{Somers2014-LithiumPMS} argue that the strong magnetic field in fast rotators during the early 
PMS enlarges the radii and diminishes the temperature of the bottom of the convective envelope, provoking over-abundances. The effective 
temperature would also be affected, due to the larger spot coverage (with cooler temperatures).
These investigations assume that the lithium spread for a given mass corresponds to real abundance differences. 
However, on the other side, \cite{Stuik1997-Pleiades-Li-K-Activityspread}, \cite{Jeffries1999-Li-Pleiades}, 
 \cite{King2000-Lithium-Rotation}, 
\cite{Barrado2001-Li-Na-Ka-Activity-Pleiades}, \cite{King2004-Activity-Alkali},  \cite{King2010-Li-K-Scatter-Pleiades},
 all in the case of the Pleiades but with very different approaches,
 have tried to verify whether the real cause
is related to the presence of surface  inhomogeneities and their effect on the observed lithium equivalent width. Some of these
works conclude that at least partially the spread is due to atmospheric effects, others argue that most come from real differences in the
depletion rate during the PMS evolution. The debate is still open. Here, we reanalyze the complete data-set using the new membership probability and rotation periods,  and try to shed new light regarding the lithium depletion in connection with rotation and stellar activity.

%%%%%%%%%%%%%%%%%%%%%%%%%%%%%%%%%%%%%%%%%%%%%%%%%%%%%%%%%%%%%%%%%%%%%%%%%%%%%%%%%%%%%%%%%%%%%  SECTION
%%%%%%%%%%%%%%%%%%%%%%%%%%%%%%%%%%%%%%%%%%%%%%%%%%%%%%%%%%%%%%%%%%%%%%%%%%%%%%%%%%%%%%%%%%%%%
%%%%%%%%%%%%%%%%%%%%%%%%%%%%%%%%%%%%%%%%%%%%%%%%%%%%%%%%%%%%%%%%%%%%%%%%%%%%%%%%%%%%%%%%%%%%%
%
\section{The data\label{sec:data}}
%
%%%%%%%%%%%%%%%%%%%%%%%%%%%%%%%%%%%%%%%%%%%%%%%%%%%%%%%%%%%%%%%%%%%%%%%%%%%%%%%%%%%%%%%%%%%%%
%%%%%%%%%%%%%%%%%%%%%%%%%%%%%%%%%%%%%%%%%%%%%%%%%%%%%%%%%%%%%%%%%%%%%%%%%%%%%%%%%%%%%%%%%%%%%
%%%%%%%%%%%%%%%%%%%%%%%%%%%%%%%%%%%%%%%%%%%%%%%%%%%%%%%%%%%%%%%%%%%%%%%%%%%%%%%%%%%%%%%%%%%%%
%%%%%%%%%%%%%%%%%%%%%%%%%%%%

\subsection{Proper motions and photometry from DANCe and TYCHO\label{sub:dance_pm}}
%
%%%%%%%%%%%%%%%%%%%%%%%%%%%%%%%%%%%%%%%%%%%%%%%%%%%%%%%%%%%%%%%%%%%%%%%%%%%%%%%%%%%%%%%%%%%%%

The starting point of this analysis is the quasi-complete census of Pleiades members 
obtained by 
\cite{Bouy2015-Pleiades},
within the DANCe project (Dynamical Analysis of Nearby Clusters, \citealp{Bouy2013-DANCE}).
The catalog has been produced by   
 essentially  retrieving all the public data corresponding to large format detectors in the available open  archives, 
reprocessing and deriving new astrometric solutions, including the correction
 of distortions due to the diverse instrumentation. 
The large temporal base-line of this amazing database and the very accurate astrometry
 have been used to derived very precise proper motions for the initial sample of almost 2,000,000 objects.

\cite{Bouy2013-DANCE} contains the  first release of the DANCe-Pleiades catalogue.
The second release is described in details in
\cite{Bouy2015-Pleiades}.
 Briefly, the main improvements with respect to the first  consists on:  the addition of
 AAVSO Photometric All-Sky Survey DR7 (APASS) $gri$ photometry and 
 the analysis of the Tycho-2 catalogue (\citealp{Hog2000.1}),
since some Pleiades members are too bright for DANCe.
 The Tycho-2 photometry was complemented with APASS, 2MASS and CMC-14 photometry, and the selection method described
 in \cite{Sarro2014.1} was applied to the merged catalogue within the same area as the DANCe-Pleiades survey. 
A total of 207 high probability, bright  members were identified, 
nicely complementing the DANCe-Pleiades catalog at the bright end of the luminosity function.

Both proper motions and the multi-wavelength photometry have fed a robust
 method based on statistic probabilities, and we have extracted the probable members, 
with probabilities larger than 75\%. This probability (PrAll, our cut-off for membership) has been derived
 using the position in several Color-Magnitude diagrams and the proper
motions. Details can be found in \cite{Sarro2014.1}.
The Pleiades list includes 2109  stellar and substellar members and 
it has a  comprehensive and homogeneous  photometry  in the Sloan and 2MASS  filters.

Note that when there is an overlap between TYCHO and DANCe,
 we have preferred the TYCHO 
 proper motions (in order to directly compare with previous works)
 and membership probabilities to DANCe results 
when plotting the data, when both data-sets were available.

% \subsection{Photometry  from DANCe\label{sub:dance_phot}}
%
%%%%%%%%%%%%%%%%%%%%%%%%%%%%%%%%%%%%%%%%%%%%%%%%%%%%%%%%%%%%%%%%%%%%%%%%%%%%%%%%%%%%%%%%%%%%%

\subsection{Lithium from the literature\label{sub:lithiumliterature}}
%
%%%%%%%%%%%%%%%%%%%%%%%%%%%%%%%%%%%%%%%%%%%%%%%%%%%%%%%%%%%%%%%%%%%%%%%%%%%%%%%%%%%%%%%%%%%%%

We have collected all the available data in the literature regarding lithium equivalent 
width --W(Li)--  in members of the Pleiades cluster. 
Namely, we have searched in the following studies:
\cite{Soderblom1993-Li-Pleiades},
\cite{GarciaLopez1994.1},
\cite{Marcy1994.1},
\cite{Basri1996.1},
\cite{Jones1996.1},
\cite{Rebolo1996.1},
\cite{Oppenheimer1997.1},
\cite{Martin1998.1},
\cite{Stauffer1998.1},
\cite{Jeffries1999-Li-Pleiades},
\cite{Martin2000.1},
\cite{Pinfield2003.1},
\cite{Margheim2007.1}, 
\cite{King2010-Li-K-Scatter-Pleiades}, and
\cite{Dahm2015-Pleiades-LDB}.
Previous works do not add additional stars to this compilation, and
have large uncertainties in their lithium equivalent widths.

In total, our compilation includes 210 objects, 
reaching around the substellar border-line at 0.072 $M_\odot$,
 and some of them have up to six individual lithium equivalent width 
measurements, totalling 398 data-points.
This collection of 201 stars will be called the Pleiades lithium sample. 
Note that there is  a gap where no observations have been executed between late K and mid-M spectral types.
Thus, W(Li) data  are not  available in the literature  
since no lithium is expected in this
spectral range --$T_{eff}$=4000-3000 K approximately--, because  this element is rapidly exhausted in PMS low-mass stars.

 The data are listed in
 Table  \ref{tab:LiCompilation} --all the lithium equivalent width for the complete data-sample,
 Table \ref{tab:TABastrometry}  --proper motions and membership  probabilities (see subsection \ref{sub:ancillaryliterature}), and
 Table \ref{tab:otherdata} --ancillary data, including effective temperatures, luminosities, final lithium equivalent widths and abundances and other data (see subsection \ref{sub:HRD} and section \ref{section:Li-Pleiades}).
 Note that for these last two tables we list only the subset
 corresponding to {\it bona fide} single stars
(additional information in subsection \ref{sub:rotation}).

\subsection{Additional ancillary data from the literature and the Virtual Observatory \label{sub:ancillaryliterature}}
%
%%%%%%%%%%%%%%%%%%%%%%%%%%%%%%%%%%%%%%%%%%%%%%%%%%%%%%%%%%%%%%%%%%%%%%%%%%%%%%%%%%%%%%%%%%%%%

For the sake of completeness, the Pleiades lithium sample also includes additional data from the literature,  namely
spectral types, 
 rotational periods,  and additional photometry in the Johnson and Cousins systems. 
The photometry comes from the Open Clusters database by Charles Prosser and John Stauffer, a careful compilation of the available data
acquired during the XX century. Table 1 from \cite{Stauffer2007.1} describes the original photometric catalogs, 
the name prefix used in each of them, the photometric bands and magnitude range for each of them.

We have compiled a secondary sample for completeness and comparison purposes,
 based on \cite{Hertzsprung1947.1}, \cite{Haro1982.1}, \cite{Hambly1993.1}, and 
 \cite{Pinfield2000.1}, corresponding to Pleiades candidate members with prefix HII, HCG, HHJ and BPL, respectively. 
After removing the duplications, we have 1131 objects. 
Then, this sample from the literature was cross-matched with the DANCe and TYCHO catalogs
in order to have homogeneous photometry in the Sloan system and membership probabilities (as derived in \citealp{Sarro2014.1}).
 We have retained only those candidate members whose 
membership probability  is larger or equal to 0.75 (as recommended in \citealp{Sarro2014.1}), 
and the sample includes 810 probable Pleiades members. 
The sample will be called the Pleiades
comparison sample.

% Pleiades lithium and Pleiades  comparison samples

% \subsection{Additional photometry from the Virtual Observatory\label{sub:ancillaryVO}}
%
%%%%%%%%%%%%%%%%%%%%%%%%%%%%%%%%%%%%%%%%%%%%%%%%%%%%%%%%%%%%%%%%%%%%%%%%%%%%%%%%%%%%%%%%%%%%%

In addition to the DANCe deep photometry (see subsection \ref{sub:dance_pm})
and the ancillary data from the literature, 
we have added public photometry using the Virtual Observatory, 
by taking advantage of  the  Virtual Observatory Sed Analyzer tool (VOSA, \citealp{Bayo2008-VOSA}, Bayo et al. 2016, submitted).
 VOSA has been designed to perform the following tasks: digests photometric data supplied by the user,  searches several photometric catalogs and theoretical models accessible through VO services,  fits all the the data to the models, computes the effective temperature and the bolometric luminosity,
and provides an estimation of the mass and age of each source.
In particular, VOSA  explores the WISE (\citealp{Cutri2012.1}), 
UKIDSS cluster (DR8, \citealp{SDSS2011}), GALEX (\citealp{Bianchi2011.1}), 
and 2MASS  (\citealp{Cutri2003.1}, \citealp{2MASS2006}) archives.
In the case of UKIDSS, the data were trimmed by selecting those magnitudes 
fainter than 13.5 (all $ZYJHK$$\le$13.5mag), in order to avoid non-linearity.
Spitzer/IRAC data from \cite{Stauffer2007.1} have also been added.

%%%%%%%%%%%%%%%%%%%%%%%%%%%%%%%%%%%%%%%%%%%%%%%%%%%%%%%%%%%%%%%%%%%%%%%%%%%%%%%%%%%%%%%%%%%%%  SECTION
%%%%%%%%%%%%%%%%%%%%%%%%%%%%%%%%%%%%%%%%%%%%%%%%%%%%%%%%%%%%%%%%%%%%%%%%%%%%%%%%%%%%%%%%%%%%%
%%%%%%%%%%%%%%%%%%%%%%%%%%%%%%%%%%%%%%%%%%%%%%%%%%%%%%%%%%%%%%%%%%%%%%%%%%%%%%%%%%%%%%%%%%%%%
%
\section{Membership\label{sec:membership}}
%
%%%%%%%%%%%%%%%%%%%%%%%%%%%%%%%%%%%%%%%%%%%%%%%%%%%%%%%%%%%%%%%%%%%%%%%%%%%%%%%%%%%%%%%%%%%%%
%%%%%%%%%%%%%%%%%%%%%%%%%%%%%%%%%%%%%%%%%%%%%%%%%%%%%%%%%%%%%%%%%%%%%%%%%%%%%%%%%%%%%%%%%%%%%
%%%%%%%%%%%%%%%%%%%%%%%%%%%%%%%%%%%%%%%%%%%%%%%%%%%%%%%%%%%%%%%%%%%%%%%%%%%%%%%%%%%%%%%%%%%%%
%%%%%%%%%%%%%%%%%%%%%%%%%%%%

%%%%%%%%%%%%%%%%%%%%%%%%%%%%%%%%%%%%%%%%%%%%%%%%%%%%%%%%%%%%%%%%%%%%%%%%%%%%%%%%%%%%%%%%%%%%%
%
%  \subsection{Position, proper motion and membership probabilities\label{sub:PM_all}}
%
%%%%%%%%%%%%%%%%%%%%%%%%%%%%%%%%%%%%%%%%%%%%%%%%%%%%%%%%%%%%%%%%%%%%%%%%%%%%%%%%%%%%%%%%%%%%%

%%%%%%%%%%%%%%%%%%%%%%%%%%%%%%%%%%%%%%%%%%%%%%%%%%%%%%%%%%%%%%%%%%%%%%%%%%%%%%%%%%%%%%%%%%%%%
%
\subsection{Proper motions and membership probabilities\label{sub:PM}}
%
%%%%%%%%%%%%%%%%%%%%%%%%%%%%%%%%%%%%%%%%%%%%%%%%%%%%%%%%%%%%%%%%%%%%%%%%%%%%%%%%%%%%%%%%%%%%%

In \cite{Bouy2013-DANCE}, we presented the DANCe
survey of the Pleiades and its new photometric and astrometric catalogue of the cluster
reaching an unprecedented accuracy of $<1$ mas/yr reaching $i\sim$24.5 mag, almost 4 mag
beyond G=20 mag, the limit provided by the Gaia mission (Prusti et al. 2016, in prep.).
Note that for the purposes of this paper, we define a star as a member
of the Pleiades if the probability of membership derived in \cite{Bouy2013-DANCE}
is larger than 75\%.

Pleiades DANCe proper motions (\citealp{Bouy2013-DANCE}) are illustrated in Fig. \ref{ProperMotion}, where
we have also included Taurus members based on their proper motions from \cite{Ducourant2005.1}.
Note the box framing the vast majority of the Taurus members and the distinct Pleiades population. However, the transition 
between the loci of both groups is smooth, and some bona-fide Pleiades members lie within the Taurus locus.
 As a matter of fact, 
there are several Pleiades candidates with low membership probability (defined as PrAll$\le$0.75)
 and detected lithium, some  are located far away from the 
Pleiades median proper motion. This sub-sample will be discussed in \ref{sub:Li_LowProbOther}.

A few lithium-rich probable members also have proper motion with significant differences from the average. However, the error-bars are large and all the photometric information suggest that they might be members.

%%%%%%%%%%%%%%%%%%%%%%%%%%%%%%%%%%%%%%%%%%%%%%%%%%%%%%%%%%%%%%%%%%%%%%%%%%%%%%%%%%%%%%%%%%%%%
%
\subsection{The Herzprung-Russell diagram\label{sub:HRD}}
%
%%%%%%%%%%%%%%%%%%%%%%%%%%%%%%%%%%%%%%%%%%%%%%%%%%%%%%%%%%%%%%%%%%%%%%%%%%%%%%%%%%%%%%%%%%%%%

We have taken advantage of VOSA (\citealp{Bayo2008-VOSA}, 
Bayo et al. 2016, submitted)
 in order  to derive basic properties
of  our samples (the  lithium and the comparison samples). 

We have derived the bolometric luminosity and effective temperature for each object, 
by using two sets of theoretical  models: 
 Kurucz (\citealp{castelli97} --any value of $T_{\rm eff}$--
and BT-Settl --$T_{\rm eff}$$\le$4400 K-- by  the Lyon group (\citealp{Allard2012.1}). 
We have restricted our computation to logg=4.5 --valid for the Pleiades age--
 and solar metallicity. 
Regarding the distance, we have selected
 133 pc (\citealp{Pinsonneault1998.1}) with  a margin of 5 pc.
A different value for the distance ($120.2\pm1.9$ pc, \citealp{vanLeeuwen2009.1})
 has been explored and its results will be discussed below.
 The non-Hipparcos values derived by trigonometric parallaxes by \cite{Soderblom2005-PleiadesDistance} or \cite{Melis2014.1}, --$134.6\pm3.1$ or $136.2\pm1.2$ pc, respectively-- are essentially the same as the one we have used.
The reddening has been fixed at  A$_v$=0.12 mag.
Note, however, that there are a dozen of Pleiades members, located South of  Merope, that have a larger reddening.
In any case, since the reddening vector goes parallel to the MS and these few members do not show
any special signature (high or low lithium content, activity), we have not taken this issue into account.
Possible  blue excesses were avoided, since neither near UV nor $u$ photometry
were not included in the fit.
Then, we have fitted a third degree polynomial to the best models, minimizing the $\chi^2$. 
Thus, we derive the optimal  $T_{\rm eff}$ and avoid the discrete values returned by VOSA.

Our derived effective temperatures have been compared with values published in the literature and with values
derived using colors and there is not obvious differences, although slow rotators (subsection \ref{sub:rotation}) 
tend to have lower $T_{\rm eff}$ when derived with VOSA, 
and fast rotators show the opposite behavior. In any case, these differences are about 100 K,
 within the uncertainties of the grid of models used by VOSA.

 Regarding the accuracy of the bolometric luminosities derived with VOSA,
 all objects have enough data-points in their
 SED to cover at least 25\% of the total bolometric flux
 (i.e., the ratio between the measured fluxes and the derived total flux, 
and the rest being  estimated with the theoretical model). More than 70\% of the Pleiades sample with measured lithium  has a ratio Flux(measured)/Flux(total) larger than 50\%.
 Thus, we can conclude that both effective temperature and bolometric luminosity are very well characterized.
This methodology does not use any estimate from bolometric corrections or a color, which
might be strongly biased. Note that individual colors might be affected by different phenomenology, such as
activity and variability, and in any case color indices cover a reduced wavelength range and a small fraction of the total luminosity,
being subject to a much larger uncertainty than our methodology, which includes all available photometry
an a very large range in wavelength coverage.

\begin{figure}
\center
\includegraphics[width=0.5\textwidth,scale=0.50]{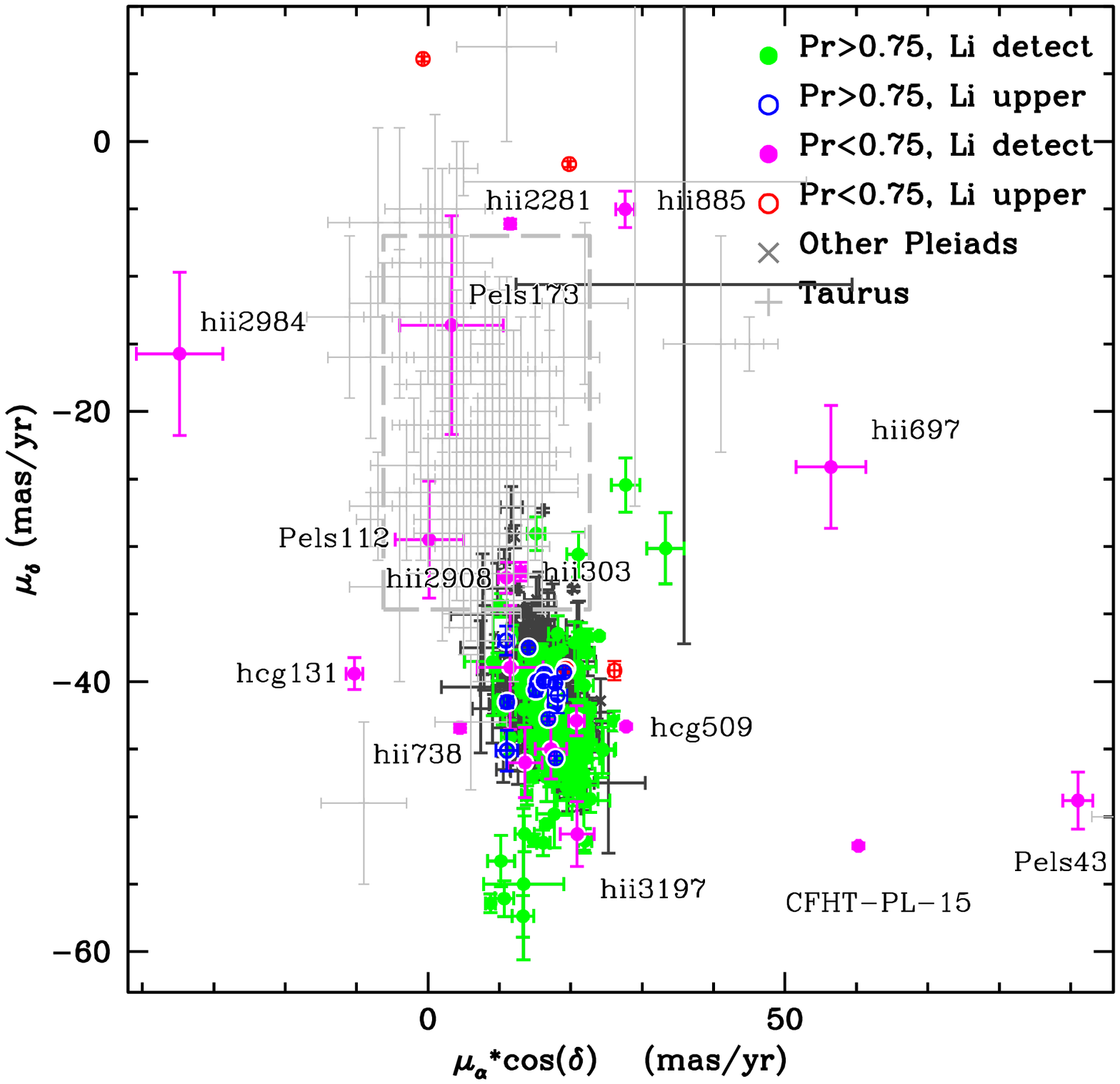} 
\caption{\label{ProperMotion} 
%, 
Proper motions for Pleiades and Taurus-Auriga members. 
 We have distinguished four cases:
 green solid circles for  lithium detection and membership probability larger than 0.75;
 blue open circles for  lithium upper limits and membership probability larger than 0.75;
 magenta solid circles for  lithium detection and membership probability less than 0.75;
 red open circles for  lithium upper limits and membership probability less than 0.75.
 Membership probabilities come from   Bouy et al. (2014).
 Light grey plus symbols are used for Taurus-Auriga members  (taken from \citealp{Ducourant2005.1}).
%   Symbols as in Fig. \ref{RA_DEC}.
}
\end{figure}

\begin{figure}
\center
\includegraphics[width=0.5\textwidth,scale=0.50]{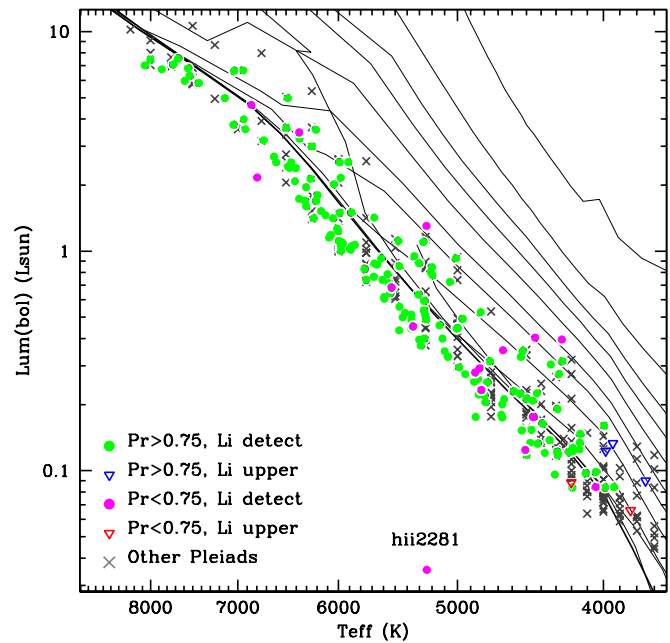} 
\caption{\label{HRDall} 
%, 
Hertzsprung-Russell diagrams displaying Pleiades data. 
The isochrones correspond to \cite{Siess2000.1} --1, 3, 5, 7, 10, 15, 20, 30, 50, 125, Myr and 5 Gyr--
and BT-Settl by  the Lyon group (\citealp{Allard2012.1}) --1, 20, 120 Myr and 10 Gyr--.
The 120/125 Myr isochrones are high-lighted.
 Symbols as in Fig. \ref{ProperMotion} .
}
\end{figure}

Our results are displayed in  Fig. \ref{HRDall}, representing a 
HR diagram (symbols as in Fig. \ref{ProperMotion}, 
but in this case only Pleiades data have been included). Two isochrone 
sets by \cite{Siess2000.1} and \cite{Chabrier2000.1} are also displayed.
Cluster members follow the 125 Myr isochrone (essentially the ZAMS 
for the spectral types F, G and K), but the width of the cluster sequence is much larger than the corresponding value
due to the binary sequence; and  an offset
 between the data and the \cite{Siess2000.1}  125 Myr isochrone is relevant. 
Note that this is also the lithium depletion boundary age of the cluster  (\citealp{Stauffer1998.2}).
This fact might be related to the distance we have used (133 pc),
 to the model itself or to the way VOSA fits the effective temperature.
 However, the proper fitting demands a distance around 115 pc,  a value even lower than the Hipparcos estimate.  Thus, the feature seems to be real and reveals a problem with the models and/or with the effect  of second order  parameters such as activity or magnetic fields.
 Other possible explanation is the  known but not completely explained 
blue excesses for Pleiades members. 
\cite{Stauffer2003.1} concludes that the color anomaly --for K members and using color-magnitude diagrams-- 
based on spottiness due to fast rotation and activity. 
In our sample, the under-luminosity respect the ZAMS also appears for a significant number of F and G stars.
 As discussed in subsection \ref{sub:rotation}, the effect should be the opposite in an HR diagram:
activity and the presence of surface inhomogeneities should shift fast rotators towards cooler temperatures
(\citealp{Jackson2013-Spots-Size}). \cite{Somers2014-LithiumPMS} (and references there) proposed inflated radii due to rotation and strong magnetic fields.
 For fainter, cooler members the BT-Settl 120 Myr isochrone describes
 very well the locus for single members. \cite{Somers2015-LithiumPleiades} achieved similar results by focusing on the Pleiades.

One object stands out in the HR diagram: HII2281.
It has an anomalous proper motion (Fig. \ref{ProperMotion} and \citealp{Hertzsprung1947.1})
and it should be demoted and rejected as a Pleiades member despite its detected lithium 
(it was already rejected by \citealp{Hertzsprung1947.1}).
 In any event, this lithium-rich star deserves further study.

%%%%%%%%%%%%%%%%%%%%%%%%%%%%%%%%%%%%%%%%%%%%%%%%%%%%%%%%%%%%%%%%%%%%%%%%%%%%%%%%%%%%%%%%%%%%%
%
\subsection{Properties of  low-probable members with  detected lithium\label{sub:Li_LowProbOther}}
%
%%%%%%%%%%%%%%%%%%%%%%%%%%%%%%%%%%%%%%%%%%%%%%%%%%%%%%%%%%%%%%%%%%%%%%%%%%%%%%%%%%%%%%%%%%%%%

There are a fair number of Pleiades members (18 without PPL-1), as defined in the literature, 
with a low probability of being 
member based on DANCe, and with lithium detection. Most of them have proper motions far away from the
cluster median (Fig. \ref{ProperMotion}, magenta solid circles). Some of them have very large errors, 
but in any case, they seem to be  interlopers, in few cases they might belong to a ``loose'' Taurus population, 
in similar fashion to what can be found in Orion (\citealp{Alves2012.1}; \citealp{Bouy2014-Orion}).

As a matter of fact, we can distinguish four different groups among these possible interlopers, depending on the
position on the proper motion diagram (Fig. \ref{ProperMotion}) and the errors, and the photometry:
The Pleiades candidates with proper motions similar to Taurus and small errors (HII2281, HII0885);
a similar group with large error-bars (HII2984, Pels173, Pels112);
another with problematic photometry and proper motions compatible with Pleiades membership
(HII0248, HII0303, HII0738, HII1275, HII1912, HII2908, HII3197, Pels41 and HCG509);  
and  the few objects with very different movement
(HII0697, HCG131, Pels43 and CFHT-PL-15).
The first two groups might belong to the the spread Taurus-like population, probably several million years old
or to the so called Local Association (the Pleiades super-cluster, \citealp{Eggen1975.1}, \citealp{Eggen1995.1}), 
a collection of young stars with similar kinematic properties, which are not necessarily coeval
 (they might be grouped by gravitational resonances of the galactic potential). This situation 
(possible relation to the Local Association) might happen for the last group. 

Regarding the nine  objects with problematic photometry, several might be binary or multiple systems
(HII0303, HII0738, HII3197, HCG509), 
due their position in the HR diagram (Fig. \ref{HRDall}). 
In addition, HCG509 and HII3197 have membership probabilities of 0.16 and 0.71, respectively, and 
even in the case of HCG509, a Li-rich mid-M (\citealp{Oppenheimer1997.1}), non-membership is not assured.
The other five stars
(HII0248, HII1275, HII1912, HII2908, and Pels41) have membership probability between 0.55 and 0.72 and
they should be considered as bona-fide members.

In any case, we can conclude that the 0.75 membership threshold is the one that gives the best compromise 
between completeness and contamination, but considering the strongly bimodality of the  distribution
 of membership probability (a huge peak at 0\% and another around 90\%)
 one can say that p=50\% is almost certainly a member, and p=15\% is also most certainly a non-member.

% %%%%%%%%%%%%%%%%%%%%%%%%%%%%%%%%%%%%%%%%%%%%%%%%%%%%%%%%%%%%%%%%%%%%%%%%%%%%%%%%%%%%%%%%%%%%%  SECTION
% %%%%%%%%%%%%%%%%%%%%%%%%%%%%%%%%%%%%%%%%%%%%%%%%%%%%%%%%%%%%%%%%%%%%%%%%%%%%%%%%%%%%%%%%%%%%
% %%%%%%%%%%%%%%%%%%%%%%%%%%%%%%%%%%%%%%%%%%%%%%%%%%%%%%%%%%%%%%%%%%%%%%%%%%%%%%%%%%%%%%%%%%%%
%
% \section{Lithium depletion  and age\label{LiDepletion}}
%
% %%%%%%%%%%%%%%%%%%%%%%%%%%%%%%%%%%%%%%%%%%%%%%%%%%%%%%%%%%%%%%%%%%%%%%%%%%%%%%%%%%%%%%%%%%%%
% %%%%%%%%%%%%%%%%%%%%%%%%%%%%%%%%%%%%%%%%%%%%%%%%%%%%%%%%%%%%%%%%%%%%%%%%%%%%%%%%%%%%%%%%%%%%
% %%%%%%%%%%%%%%%%%%%%%%%%%%%%%%%%%%%%%%%%%%%%%%%%%%%%%%%%%%%%%%%%%%%%%%%%%%%%%%%%%%%%%%%%%%%%
% %%%%%%%%%%%%%%%%%%%%%%%%%%%

% %%%%%%%%%%%%%%%%%%%%%%%%%%%%%%%%%%%%%%%%%%%%%%%%%%%%%%%%%%%%%%%%%%%%%%%%%%%%%%%%%%%%%%%%%%%%
% 
% \subsection{The lithium abyss\label{sub:LithiumAbyss}}
% 
% %%%%%%%%%%%%%%%%%%%%%%%%%%%%%%%%%%%%%%%%%%%%%%%%%%%%%%%%%%%%%%%%%%%%%%%%%%%%%%%%%%%%%%%%%%%%

%%%%%%%%%%%%%%%%%%%%%%%%%%%%%%%%%%%%%%%%%%%%%%%%%%%%%%%%%%%%%%%%%%%%%%%%%%%%%%%%%%%%%%%%%%%%%
%%%%%%%%%%%%%%%%%%%%%%%%%%%%%%%%%%%%%%%%%%%%%%%%%%%%%%%%%%%%%%%%%%%%%%%%%%%%%%%%%%%%%%%%%%%%%
%%%%%%%%%%%%%%%%%%%%%%%%%%%%%%%%%%%%%%%%%%%%%%%%%%%%%%%%%%%%%%%%%%%%%%%%%%%%%%%%%%%%%%%%%%%%%
%
% \subsection{Lithium abundance and its dependence with the stellar properties and evolution:  FGK members\label{sub:ALi_FGK}}
\section{Lithium in members of the Pleiades\label{section:Li-Pleiades}}
%
%%%%%%%%%%%%%%%%%%%%%%%%%%%%%%%%%%%%%%%%%%%%%%%%%%%%%%%%%%%%%%%%%%%%%%%%%%%%%%%%%%%%%%%%%%%%%
%%%%%%%%%%%%%%%%%%%%%%%%%%%%%%%%%%%%%%%%%%%%%%%%%%%%%%%%%%%%%%%%%%%%%%%%%%%%%%%%%%%%%%%%%%%%%
%%%%%%%%%%%%%%%%%%%%%%%%%%%%%%%%%%%%%%%%%%%%%%%%%%%%%%%%%%%%%%%%%%%%%%%%%%%%%%%%%%%%%%%%%%%%%

%%%%%%%%%%%%%%%%%%%%%%%%%%%%%%%%%%%%%%%%%%%%%%%%%%%%%%%%%%%%%%%%%%%%%%%%%%%%%%%%%%%%%%%%%%%%%
%
\subsection{Building a clean  a lithium sample for the Pleiades\label{subsub:Li-sample}}
%
%%%%%%%%%%%%%%%%%%%%%%%%%%%%%%%%%%%%%%%%%%%%%%%%%%%%%%%%%%%%%%%%%%%%%%%%%%%%%%%%%%%%%%%%%%%%%

In order to revisit the lithium problem,
we have created an homogeneous sample (at least as much as possible) of bona-fide Pleiades members.
First, we have selected only those with membership probabilities larger than 0.75 (see \ref{sub:Li_LowProbOther}).
Then, for those with several measured W(Li), we have selected only one value among our complete data set of published W(Li).
To do so we have followed two different approaches: the primary --selection A-- is based on the spectral resolution,
whereas the secondary --selection B-- essentially depends on the S/N.

In the first case the rationale responds to the clear separation from the lithium feature at 6707.8 \AA{} of the
contaminant iron line at 6707.4 \AA --i.e., both can be measured.
The prioritization has been:
\cite{King2010-Li-K-Scatter-Pleiades}  --R$\sim$60\,000 in a 9.2m telescope, 
\cite{Jones1996.1} --R$\sim$45\,000 in a 10m telescope,
\cite{Oppenheimer1997.1} --same as previous,
\cite{Soderblom1993-Li-Pleiades} --R$\sim$50\,000 in a 3m telescope,
\cite{Margheim2007.1} --R$\sim$12\,800 in a 3.5m telescope with multiplexing, 
\cite{Jeffries1999-Li-Pleiades} --R$\sim$14\,500 in a 2.5m telescope, and
\cite{GarciaLopez1994.1} --R$\sim$10\,000 in a 4.2m telescope.

For selection B, where the achieved signal-to-noise ratio is paramount, so we have preferred
 the data coming from spectra acquired with  10-m class telescopes or
4-m class with large multiplexing (i.e., large exposure times) over conventional instruments
 (long slit, echelle spectrographs) in 4-m, 3-m or 2-m class telescopes.
In this case we have followed the  hierarchy:
\cite{King2010-Li-K-Scatter-Pleiades}, 
\cite{Margheim2007.1}, 
\cite{Jones1996.1},
\cite{Jeffries1999-Li-Pleiades},
\cite{Oppenheimer1997.1},
\cite{GarciaLopez1994.1}, and
\cite{Soderblom1993-Li-Pleiades}.
 Thus, we have selected the values with the smallest individual errors in W(Li), while
 preserving some homogeneity in the data-set.

  The most significant difference between both selections, due to the sample sizes, correspond to values from \cite{Margheim2007.1} and \cite{Soderblom1993-Li-Pleiades}. The first work compared both data-sets and derived a systematic difference with a mean value of 12 m\AA, being the most recent, taken at higher S/N and lower spectral resolution, higher. This W(Li) offset does not depend in the specifics value and it is essentially constant for all equivalent widths. \cite{Margheim2007.1} ascribed this difference to a lower continuum for low S/N spectra,  although the net effect of spectra taking at significant resolution is to decrease the measured equivalent width. However, \cite{Margheim2007.1} supports his conclusion with a comparison between abundances derived with equivalent widths and with  spectral synthesis.

 Regardless the offsets in the individual sources of lithium equivalent widths, our analysis has been done in parallel with both selections and we have not found any significant difference. All values are listed in Table \ref{tab:otherdata}.

%%%%%%%%%%%%%%%%%%%%%%%%%%%%%%%%%%%%%%%%%%%%%%%%%%%%%%%%%%%%%%%%%%%%%%%%%%%%%%%%%%%%%%%%%%%%%
%
\subsection{Lithium in FGK members\label{section:Li-FGK}}
%
%%%%%%%%%%%%%%%%%%%%%%%%%%%%%%%%%%%%%%%%%%%%%%%%%%%%%%%%%%%%%%%%%%%%%%%%%%%%%%%%%%%%%%%%%%%%%

\begin{figure}
\center
\includegraphics[width=0.5\textwidth,scale=0.50]{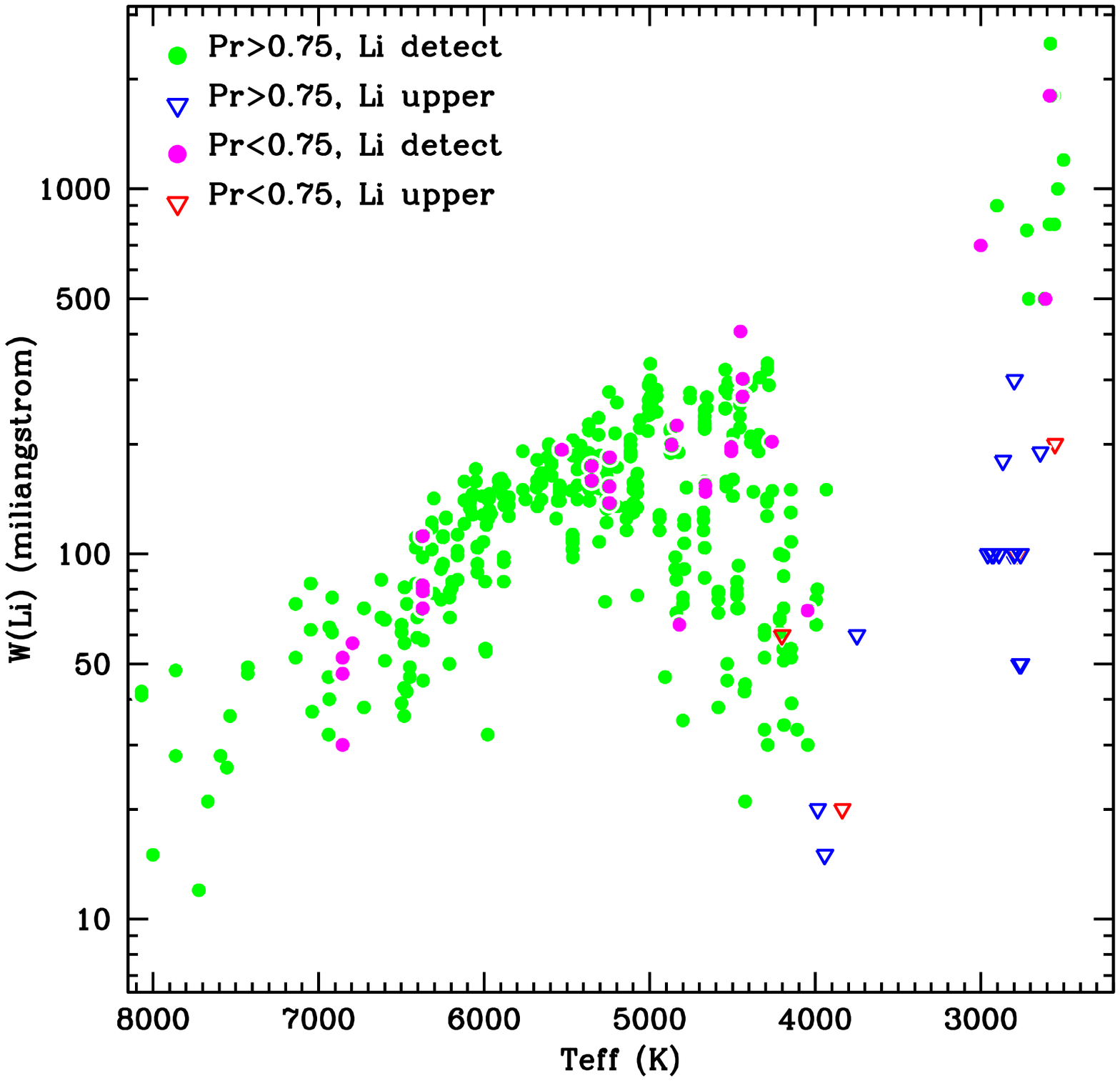} 
\includegraphics[width=0.5\textwidth,scale=0.50]{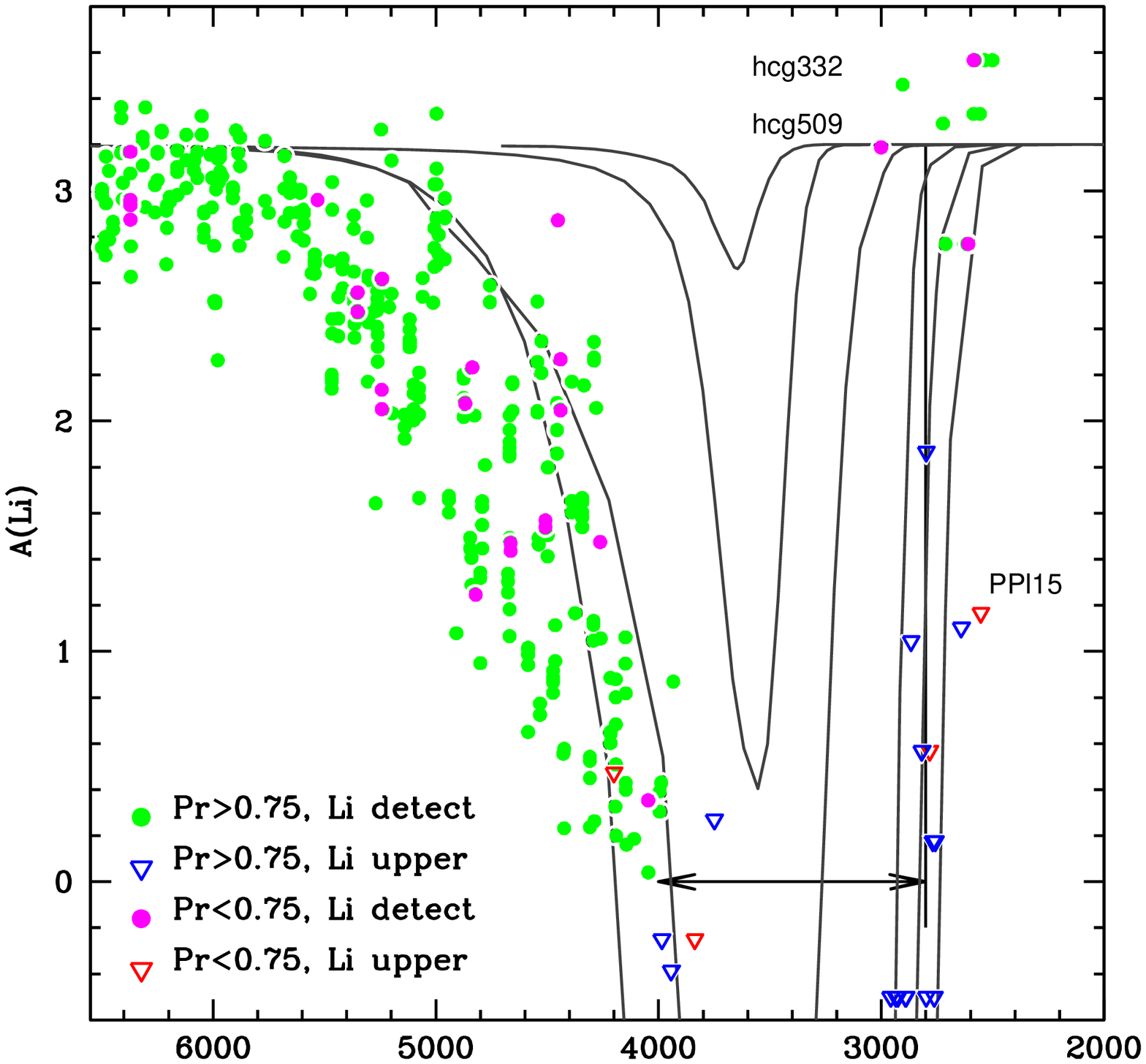} 
\caption{\label{Teff_WLi_and_ALi_LDB} 
%, 
Lithium equivalent width (top) and lithium abundance (bottom) versus effective temperature.
The diagram shows all relevant  data, including multiple measurements for the same stars  and known binaries. 
Symbols correspond to:
green solid circles  are used for  lithium detection and membership probability larger than 0.75 (from Bouy et al. 2014);
blue open triangles for  lithium upper limits and membership probability larger than 0.75;
magenta solid circles for  lithium detection and membership probability less than 0.75;
red open triangles for  lithium upper limits and membership probability less than 0.75.
Membership probabilities come from  (from Bouy et al. 2014).
The curves correspond to BT-Settl models from \cite{Allard2012.1}. The central dips are for 10 and 20 Myr, 
the blue side of the lithium abyss includes values for 50 and 125 Myr, and the red side displays the 
computation for 50, 90, 125 and 150 Myr.
}
\end{figure}

Figure \ref{Teff_WLi_and_ALi_LDB} --top panel-- shows the lithium equivalent width versus the effective temperature.
We have not plotted the errors in W(Li) for clarity, but they are not large in most of the cases
(see Table \ref{tab:otherdata}).
The large spread for a given temperature is quite obvious, as is the dramatic drop after $T_{\rm eff}$=4400 K  and the 
location of the LDB at 2800 K  (the blue and the red sides of the lithium abyss, respectively).

We have derived the lithium abundances, defined as A(Li)=12+log (Li/H),
using  curves of growth by \cite{Soderblom1993-Li-Pleiades}.
For objects with effective temperature cooler than 4000 K 
(the coolest of the curves from \citealp{Soderblom1993-Li-Pleiades}), we have assumed
this temperature, and derived an abundance just as a reference. 
We have avoided the mixing of several sets of curves of growth because, for very low temperatures, 
they produce very different 
values (see, for instance,  the comparison between \citealp{Palla2007.1} and \citealp{Pavlenko1996.1}).
In any case, this assumption does not affect our discussion,
 since most of the object below 4000 K have lithium upper limits and those with actual detections 
have $T_{\rm eff}$$\le$2600 and should have a cosmic, undepleted abundance (i.e., A(Li)=3.1--3.2).

In order to estimate the errors in the abundance, we have taken into account the original errors in W(Li), as taken from the literature, and the estimated errors for the $T_{\rm eff}$ fit by bootstrapping their values inside their uncertainties.

The results have been displayed in Fig. \ref{Teff_WLi_and_ALi_LDB} --bottom panel. 
The diagram also includes the theoretical lithium depletion computed in the BT-Settl models (50, 90, 125, and 150 Myr). As can be seen, 
the models reproduce very well the location of the LDB, but since they do not include all the complex affects going on for F, G and K stars, 
they are only indicative for the blue side.

The lithium spread for stars with the same temperature is very evident for any spectral type,
 but it is even more conspicuous for $T_{\rm eff}$=5200-4000 K. In fact, there is a relation with the bolometric luminosities:
 the higher the W(Li), the higher the bolometric luminosity and might be related with binarity 
(see the case of the Hyades,  \citealp{Barrado1996-Li-Hyades}) and/or rotation. This will be  examined in greater detail in 
subsection \ref{subsub:Li_GKspread}.

Some stars with high lithium abundance and $T_{\rm eff}$ around 4500 K have low membership probability.
 They tend to have larger luminosities and some of them might be pollutants (see subsection \ref{sub:Li_LowProbOther}).

The  Pleiades members confirmed by DANCe with a G-K spectral type have  lithium abundances 
corresponding, {\it grosso modo}, to the expected values   for their age, 
even taking into account the lithium spread for the same $T_{\rm eff}$.
This fact does not agree with the models of \cite{Baraffe2010.1}, where they
predict that episodic accretion events during the PMS could produce ZAMS age GK stars 
that are severely depleted in lithium (hence calling into question use of lithium as
a membership indicator in young clusters like the Pleiades).
A similar investigation by \cite{Sergison2013.1} has not found this effect either.

As stated before, the lithium depletion pattern has received a significant amount of attention both from
 the theoretical and the observational points of view.
Very recently \cite{Somers2014-LithiumPMS} linked  the effect of inflated radii due to rotation and strong magnetic fields with  lithium depletion.
Before, 
 \cite{Bouvier2008.1} postulated that the basic reason was different disk life-time for fast and slow rotators, the rotation evolution,
 and the extra lithium mixing due to the shear at the bottom of the convective zone. 
However, it is not clear that such models can explain the full range in lithium
seen in the Pleiades: 
\cite{Eggenberger2012-Li-Rotation-Disk-PMS} developed the disk life-time concept further for a 1 $M_\odot$ model 
and predicted a maximum of 0.25 dex change in abundance, totally insufficient.
The lack of depletion for 0.9  $M_\odot$ at 5000 K for some Pleiades members, normally fast rotators, is unexplained by any model,
and in any case, regarding the results by \cite{Somers2014-LithiumPMS},
the  lithium-rotation connection  cannot be reproduced by their scenario, since essentially  all very young PMS stars are 
in a saturated activity regime, and there is thus no activity (hence radius inflation) -- rotation connection
(see the case of  h Per, \citealp{Argiroffi2013.1}).

%%%%%%%%%%%%%%%%%%%%%%%%%%%%%%%%%%%%%%%%%%%%%%%%%%%%%%%%%%%%%%%%%%%%%%%%%%%%%%%%%%%%%%%%%%%%%
%
\subsection{Binarity and Rotation: an effect on the HRD location ?\label{sub:rotation}}
%
%%%%%%%%%%%%%%%%%%%%%%%%%%%%%%%%%%%%%%%%%%%%%%%%%%%%%%%%%%%%%%%%%%%%%%%%%%%%%%%%%%%%%%%%%%%%%

\begin{figure}
\center
\includegraphics[width=0.5\textwidth,scale=0.50]{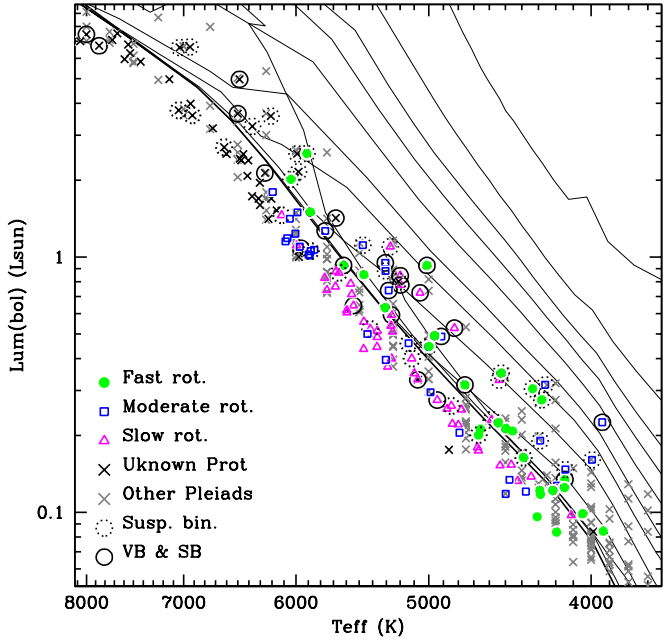} 
\caption{\label{HRDbinaries} 
%, 
Zoom for the Hertzsprung-Russell diagram displaying Pleiades FGK members.
The isochrones correspond to \cite{Siess2000.1}, with ages of 1, 3, 5, 7, 10, 15, 20, 30, 50, 125, Myr and 5 Gyr.
The 125 Myr isochrone is high-lighted.
% In addition to the symbols used in  Fig. \ref{Teff_WLi_and_ALi_LDB}, we have included 
Fast (green solid circles), moderate (open blue squares) and slow  (open magenta triangles) rotators 
are defined by their location within the
period vs. luminosity plot shown in Fig. \ref{Prot_LumBol}. 
The same symbol and color code are used in the and following figures.
Empty
large black circles for binaries (broken lines for suspected binaries). 
}
\end{figure}

The evolution of the lithium abundance in a star is affected by rotation. Binarity might play a role too.
 This fact is very relevant in the case of Tidally Locked 
Binary Systems (with Prot$\le$15 days), as shown in chromospherically active binaries
 (\citealp{Barrado1997-Age-Dwarfs-RSCVn}),  Hyades  (\citealp{Barrado1996-Li-Hyades}) or Praesepe members (\citealp{King1996-Li-Praesepe-TLBS}).
Several works have searched for binaries in the Pleiades 
(\citealp{Mermilliod1992_PleiadesBinarity};  \citealp{Rosvick1992_PleiadesRV};  
\citealp{Bouvier1997.Pleiades_binaries};  \citealp{Queloz1998_PleiadesVrot})
 and we have identified them in our sample.
 However,  some unaccounted for binary and multiple systems might be present too. 
In order to detect them, we have investigated the position of our sample of G and K 
Pleiades in an HR diagram. 
Known visual and spectroscopic binaries have been highlighted with large black circles in  Fig. \ref{HRDbinaries}.
They tend to be brighter than single stars for the same effective temperature, 
 although there is a significant number of bright,  ``single'' stars
(indicated as broken large black circle in the diagram). 
Therefore, we have classified them  as ``suspected'' binaries.
The possible relation between rotation and the lithium spread, including binaries, 
 will be investigated in subsection \ref{subsub:Li_GKspread}.

\begin{figure}
\center
\includegraphics[width=0.5\textwidth,scale=0.50]{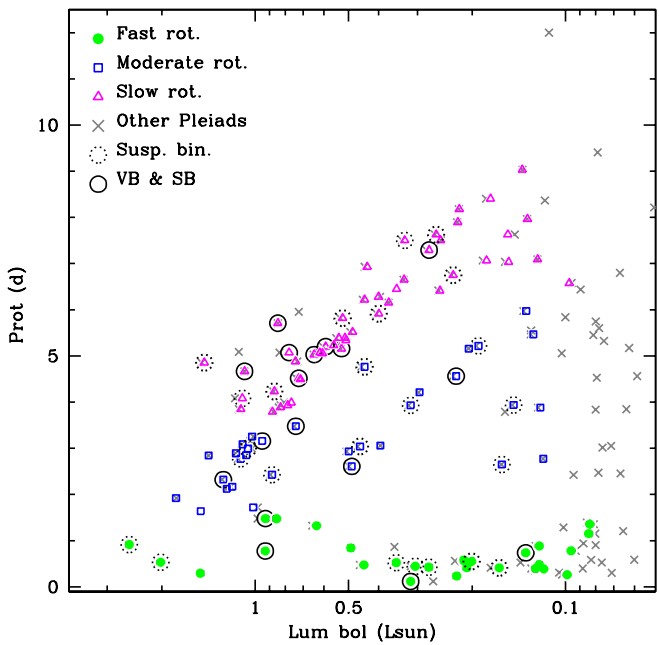} 
\caption{\label{Prot_LumBol} 
%, 
Rotation period versus the bolometric luminosity.
Symbols as in Fig. \ref{HRDbinaries}.
}
\end{figure}

Regarding rotation, we have cross-correlated our data with the rotational periods derived  by 
  \cite{Hartman2010.1}, and we have classified our sample of lithium members in
 fast, moderate and slow rotators based on its distribution for different bolometric luminosities 
(see Fig. \ref{Prot_LumBol})
or effective temperatures (the results are essentially the same).
 Although our criteria are somehow arbitrary (the borderlines between groups, see the discussion by
\citealp{Bouvier1997.1} regarding the rotation dichotomy  and the link with circumstellar disks), they
do not affect the general conclusions. Note, however, that rotational periods are not always accurate.
As a matter of fact, the comparison between the values from   \cite{Hartman2010.1} and 
preliminary estimates from Kepler K2 (\citealp{Howell2014-Kepler-K2}) indicates that HII1095 belongs 
to the slow rotation group.

It is worth noting that known binaries tend to rotate slower than single stars 
(again from Fig. \ref{Prot_LumBol}), 
at least for solar-type stars, where the statistics are more robust.
The explanation might be due to the synchronization effect: a fraction of the Pleiades 
binaries would have orbital periods between 2 and 7 days,
 forcing the rotation period to be equal to the orbital one,
 and larger than the corresponding value for single, fast rotators with the same 
luminosity. Note, however, that there is a bias for finding slowly rotating spectroscopic binaries, since short period binaries can be confused with rapid rotators due to the blending of the lines.

Moreover,  the symbol code used on Fig. \ref{HRDbinaries} also includes information regarding rotation. 
An interesting effect might take place: single stars with  fast rotation might be, on average, brighter or cooler than
other slower single members.
However, the effect is far from proved and a Color-Magnitude Diagram shows that single, fast rotating stars fall 
on top of the slower ones (\citealp{Stauffer1987-rotation-Pleiades}).
Lately, \cite{Jackson2013-Spots-Size} and \cite{Jackson2014-Spots-Radii-PMS} 
have investigated this issue and concluded that active Pre-Main Sequence stars
have an expanded radius due to the presence of spots, slowing down the descent along  the Hayashi tracks, although they cannot
rule out that the strong magnetic fields are  inhibiting the  convective flux
transport.
\cite{Somers2014-LithiumPMS} predicts larger radius (and cooler temperatures) for fast rotators.
A different evolutionary history for fast rotators should have an effect on the lithium content, 
which would be maintained over the whole stellar time-span.
However, the  spread for single Hyades stars at about 600 Myr  is much smaller than the observed range 
for the Pleiades.
 If the effect is real, our interpretation is that the fast rotators appear cooler due to activity, and the overall
bolometric luminosity, a quantity that depends on the nuclear reaction in the stellar core, remains essentially unaffected.
A significant fraction of their surface should be covered with solar-like spots, with lower temperature, and the Spectral Energy
Distribution should be affected by the presence of the spot, shifting the distribution to the red and keeping the total energy output.
Note that in order to compute  $Lum(bol)$ or $T_{\rm eff}$ we do not rely on bolometric corrections and colors, which can be affected by activity and 
the presence of surface inhomogeneities.
However, we cannot completely discard  the other effect discussed above or changes
 on the stellar structure due to rotation and the centrifugal forces.
Color effects due to rotation and activity
will be studied in depth somewhere else.

%%%%%%%%%%%%%%%%%%%%%%%%%%%%%%%%%%%%%%%%%%%%%%%%%%%%%%%%%%%%%%%%%%%%%%%%%%%%%%%%%%%%%%%%%%%%%
%
\subsection{The lithium spread for G and K stars: rotation versus activity\label{subsub:Li_GKspread}}
%
%%%%%%%%%%%%%%%%%%%%%%%%%%%%%%%%%%%%%%%%%%%%%%%%%%%%%%%%%%%%%%%%%%%%%%%%%%%%%%%%%%%%%%%%%%%%%

The lithium spread, specially for K stars, has been discussed by
\cite{Soderblom1993-Li-Pleiades},
and extensively examined in the literature for the Pleiades
(\citealp{GarciaLopez1994.1}, \citealp{Jones1996.1}, \citealp{Oppenheimer1997.1}, 
\citealp{Jeffries1999-Li-Pleiades}, 
and \citealp{King2010-Li-K-Scatter-Pleiades}) and for other clusters such as  NGC2451 (\citealp{Hunsch2004.1}).
For more details, see the introduction.

That lithium equivalent widths for young stars are enhanced for
rapidly rotating and active K dwarfs is well established, but 
there is still no universally accepted physical explanation nor
agreement as to whether the observed effect is due to differences
in lithium abundance or instead is a consequence of differing
line formation processes.
We have already discussed its relation with activity via surface inhomogeneities in
\cite{Barrado2001-Li-Na-Ka-Activity-Pleiades}, and compared with other lines produced by alkali elements such as sodium and potassium,
 which are not depleted in the stellar interior, 
 concluding that at least {\it partially} the change in effective temperature (lower when compared with the spotless photosphere) 
should produce an increase in the observed lithium equivalent width for the same abundance.
A similar conclusion has been derived by \cite{Randich2001-Li-K-IC2602}, 
at least for stars warmer than $\sim5400$ K: 
activity modifies the potassium equivalent widths, and thus
the inferred potassium abundances (but see also the study of stars
in the Alpha Per cluster by \citealp{Martin2005.1}).

\begin{figure*}
\center
\includegraphics[width=0.45\textwidth,scale=0.50]{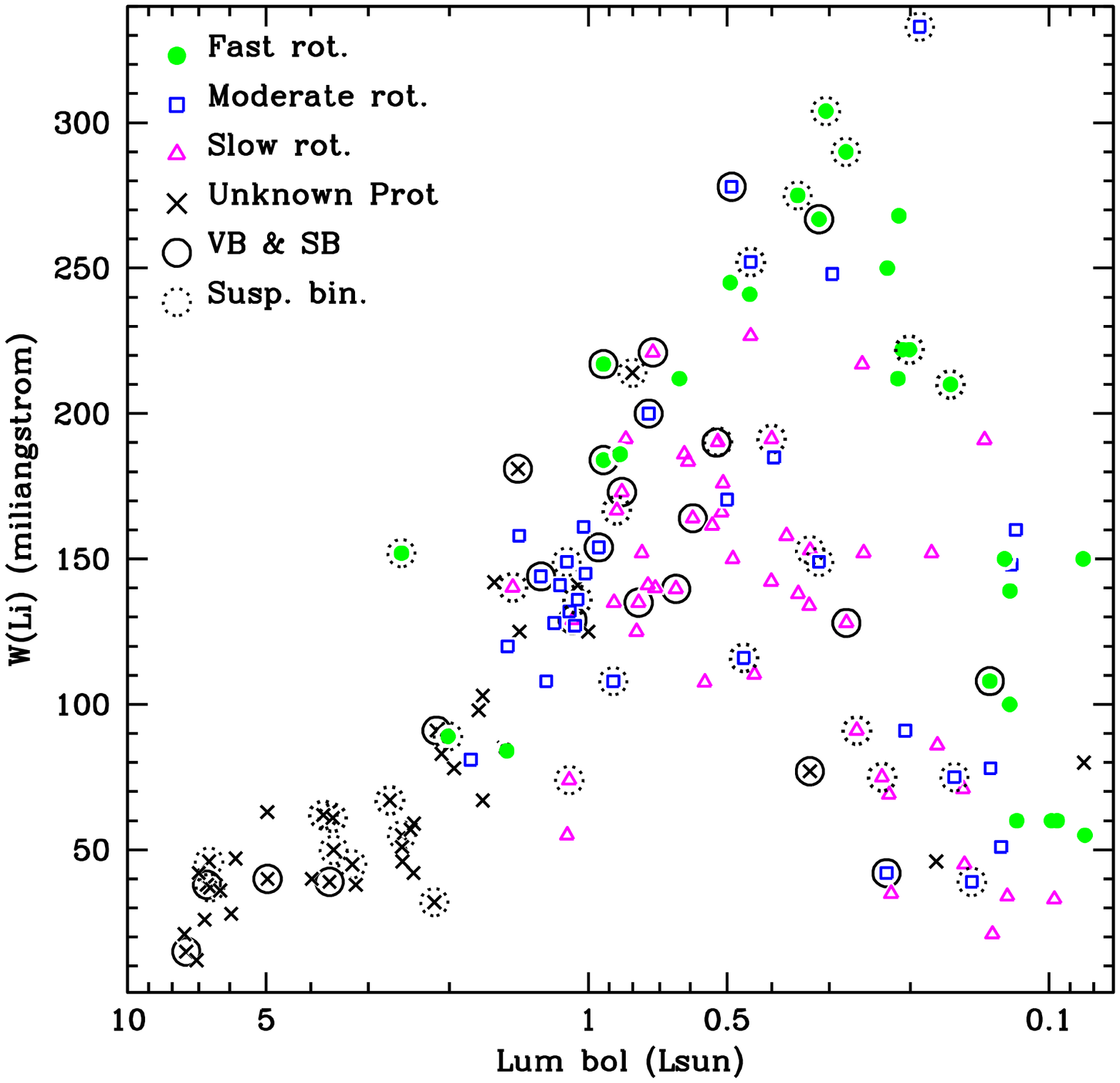} 
\includegraphics[width=0.45\textwidth,scale=0.50]{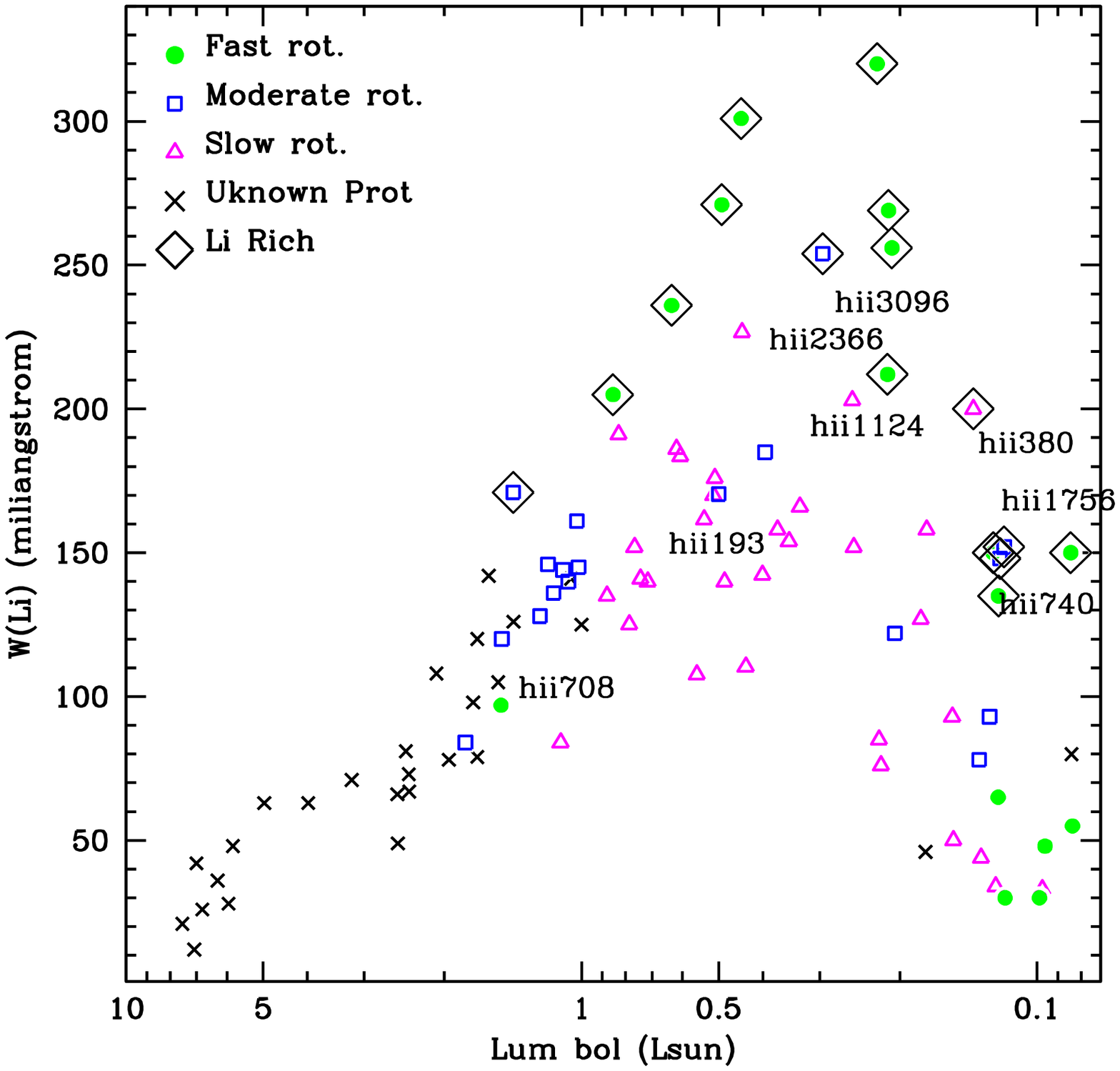}
\includegraphics[width=0.45\textwidth,scale=0.50]{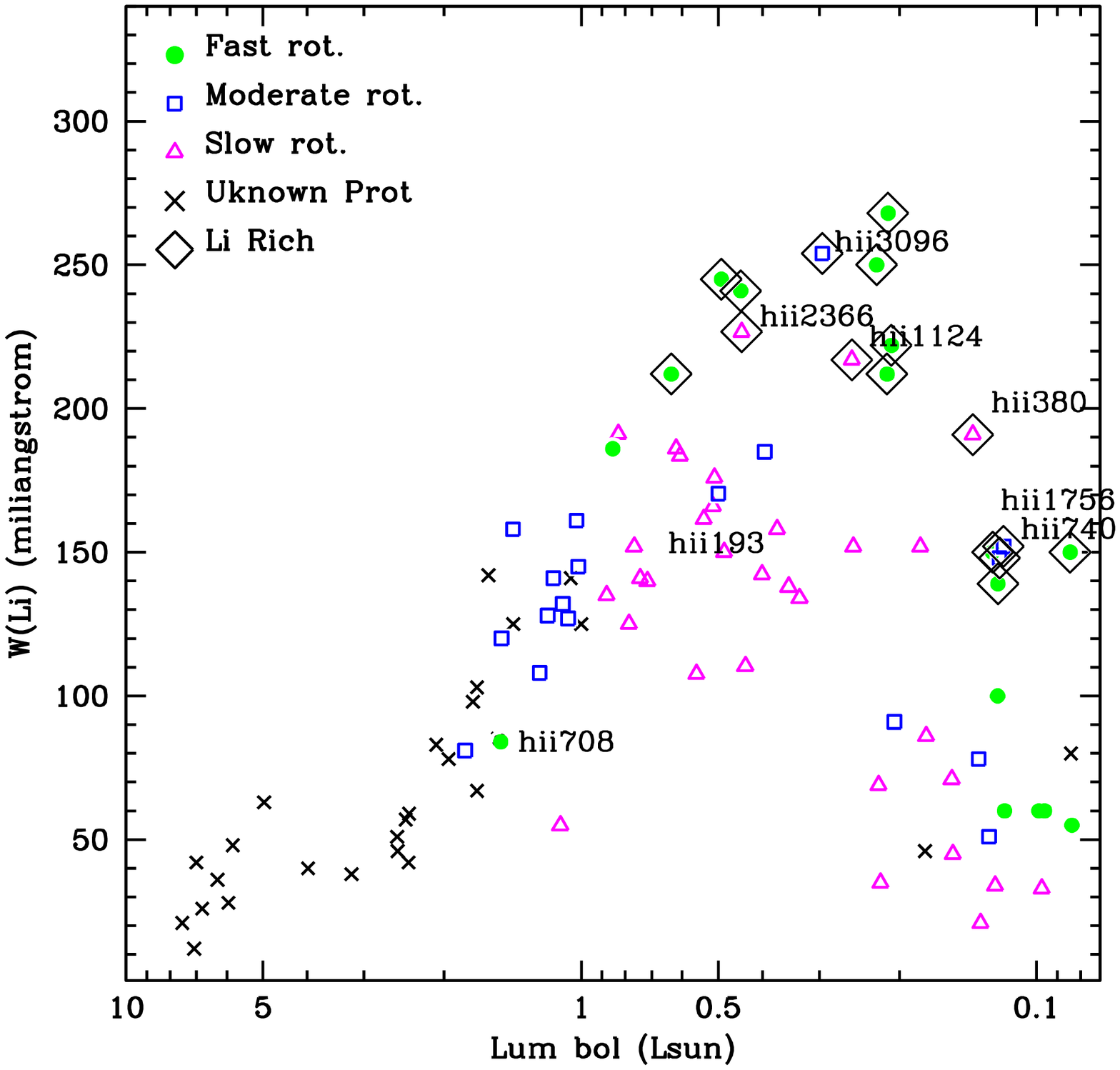}
\caption{\label{LumBol_WLi} 
%, 
 Lithium equivalent with versus the bolometric luminosity
for the high probable members (probability larger than 0.75).
For those stars with multiple values of W(Li), only our selection is represented (see text).
{\bf Top: } Binaries and suspected binaries
 are included and highlighted with large open circles.
{\bf Middle and bottom: } Only single stars are displayed. Lithium-rich stars are also indicates with
large, open diamonds. In the first case we have used the W(Li) selection based on the S/N --selection B-- whereas in the second
panel  we display the data based on spectral resolution --selection A.
}
\end{figure*}

Figure \ref{LumBol_WLi} displays our selected values of lithium equivalent width versus
 the bolometric luminosity (in order to avoid the possible effect by the 
activity/spotness on the effective temperature). The top panel includes single, 
known and suspected binaries, whereas, for the sake of simplicity, the bottom
panel only displays single stars. Colors and symbols differentiate the three classes of rotators (see Fig. \ref{Prot_LumBol}).
The data for single stars appear in Table \ref{tab:TABastrometry} and Table \ref{tab:otherdata}.  
 The  relation between the equivalent width and the
rotation for single stars, for the same luminosity, is quite striking, specially for
 $Lum(bol)=0.5-0.2$ $L_\odot$ (about K2-K7 spectral types  or 0.75--0.95 $M_\sun$), but even reaching
1.0 $L_\odot$ (about G5 or 1.1 $M_\sun$).
Stars with very large lithium equivalent width are marked with big open diamonds and
will be discussed in some detail.  In the case of binaries,
fast rotating systems (or suspected binaries) have much larger equivalent widths than slow rotators.
Is this a consequence of rotation per se?
Note that
HII3096 and HII380, 
the first one having intermediate rotation rate and the last being a slow rotator,
have lithium equivalent widths larger than the corresponding
 value for similar stars.
In order to explain these anomalies,  lets have a look at the activity effect on W(Li) via surface inhomogeneities.

\begin{figure*}
\center
\includegraphics[width=0.45\textwidth,scale=0.50]{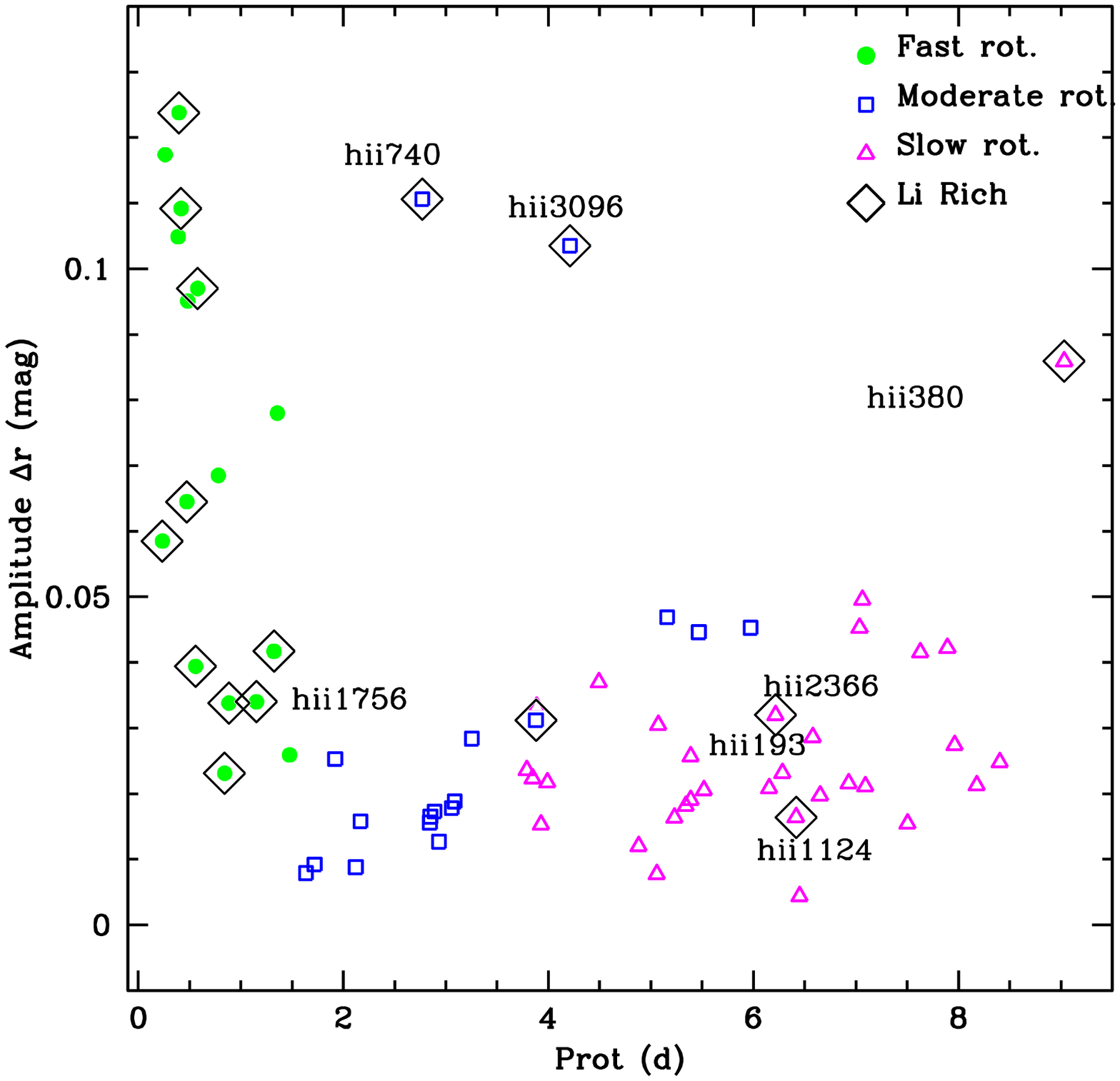} 
\includegraphics[width=0.45\textwidth,scale=0.50]{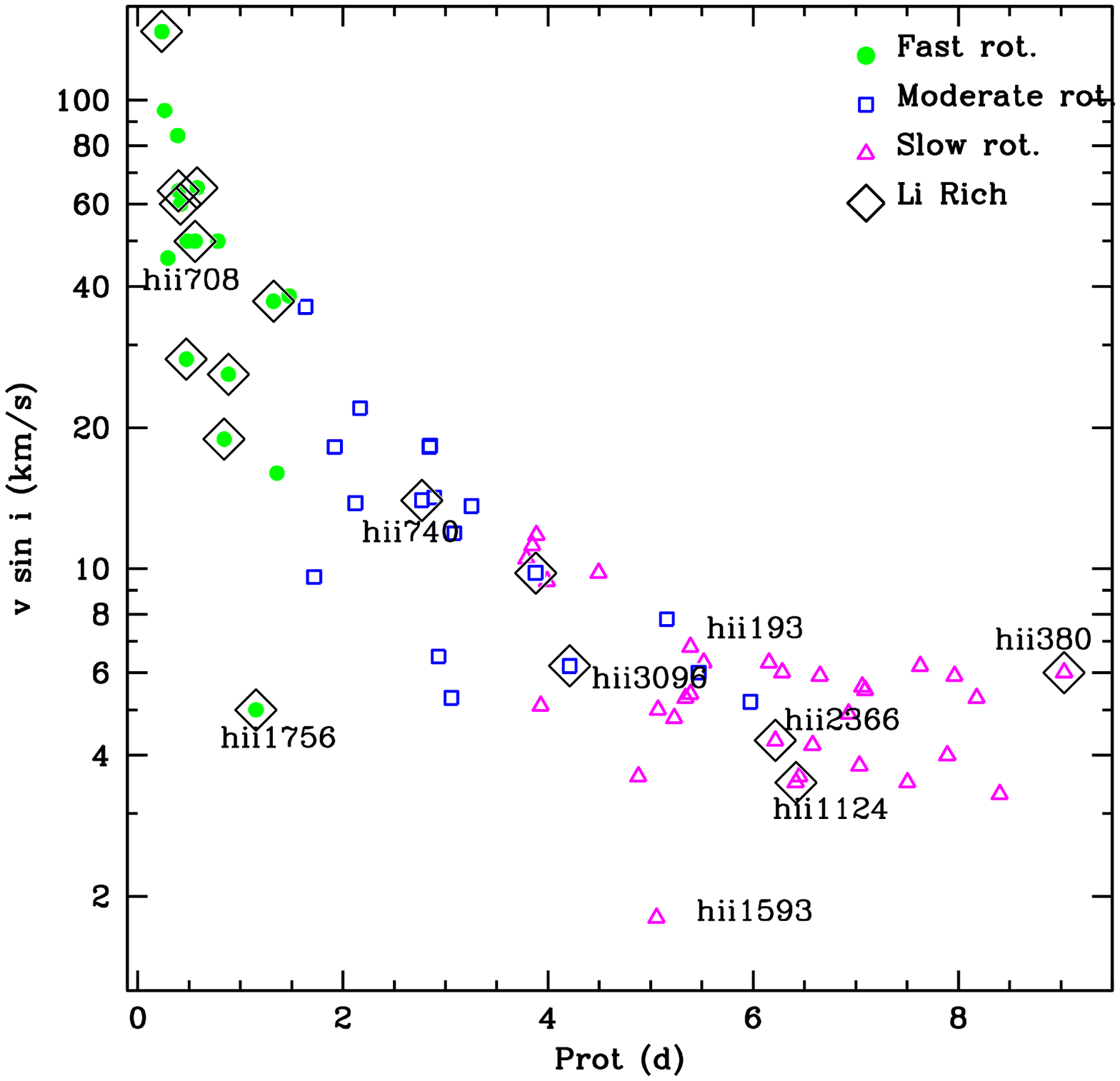} 
\includegraphics[width=0.45\textwidth,scale=0.50]{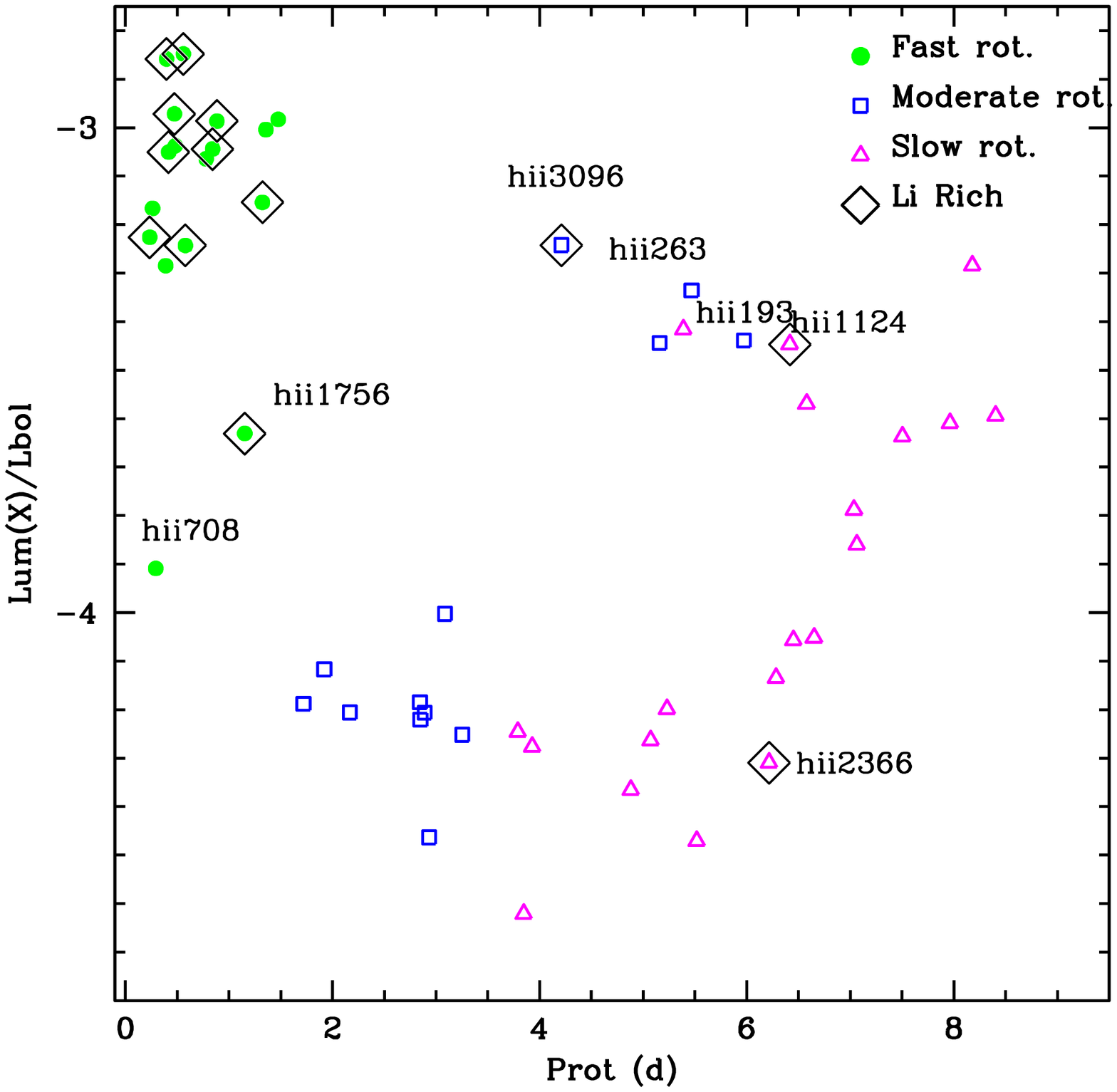} 
\caption{\label{Prot_Ampl_Vsini} 
%, 
Photometric variability in the $r$ band ({\bf top}).
  the projected equatorial velocity ({\bf middle}),
 and the fraction between the Lx and the bolometric luminosities ({\bf bottom})
 versus the rotational period. 
 Only single stars are displayed. The lithium rich stars come from selection A, based on spectral resolution,
 although the result is practically identical if the stress is put on the S/N.
Note that HII380 and HII740 do not have a measured value of Lx, and HII708 does not have a measured value of the
photometric amplitude,  and therefore they are not
represented on some of these diagrams.
 }
\end{figure*}

Figure \ref{Prot_Ampl_Vsini}a shows the relation between the photometric variability (the amplitude of the light curve) and 
the rotational period. As expected, slow rotators have a reduced variability.
On the contrary, fast rotators display a large range for the photometric amplitude.
 There are two explanations for this: 
either the spots might be distributed very evenly on the stellar surface,
or the inclination angle is large (near pole-on), so a stellarspot or a group of them are visible during a significant fraction of the period.
Therefore, even if the spot filling factor is large, the photometric variability should be small.
In fact, this is what seems to be happening for HII1756 --fast rotator-- and HII1593 --slow rotator, 
as shown in Fig. \ref{Prot_Ampl_Vsini}b, where we display the $vsini$ versus the rotational period. 
In any case, some might have genuine low activity (perhaps reassembling a solar minimum or even a Maunder minimum).
Several Pleiades members have been labeled in the diagram. All four slow and moderate rotator with large variability 
have large W(Li). 

What about other signposts of activity?
We have compared the X-ray emission (the ratio between the X-ray and the bolometric luminosities) with the rotational periods 
(Fig. \ref{Prot_Ampl_Vsini}c).
As expected, the large majority of fast rotators (short $Prot$) have Lum $L_{X}$/$L_{Bol}$ near the saturation at a ratio
close to $-3$. Most of these Pleiades members display, indeed, a lithium excess, even if the $Prot$ is not remarkably short.
In any case, the figures show a trend, no a clear-cut relationship.

\begin{figure}
\center
\includegraphics[width=0.5\textwidth,scale=0.50]{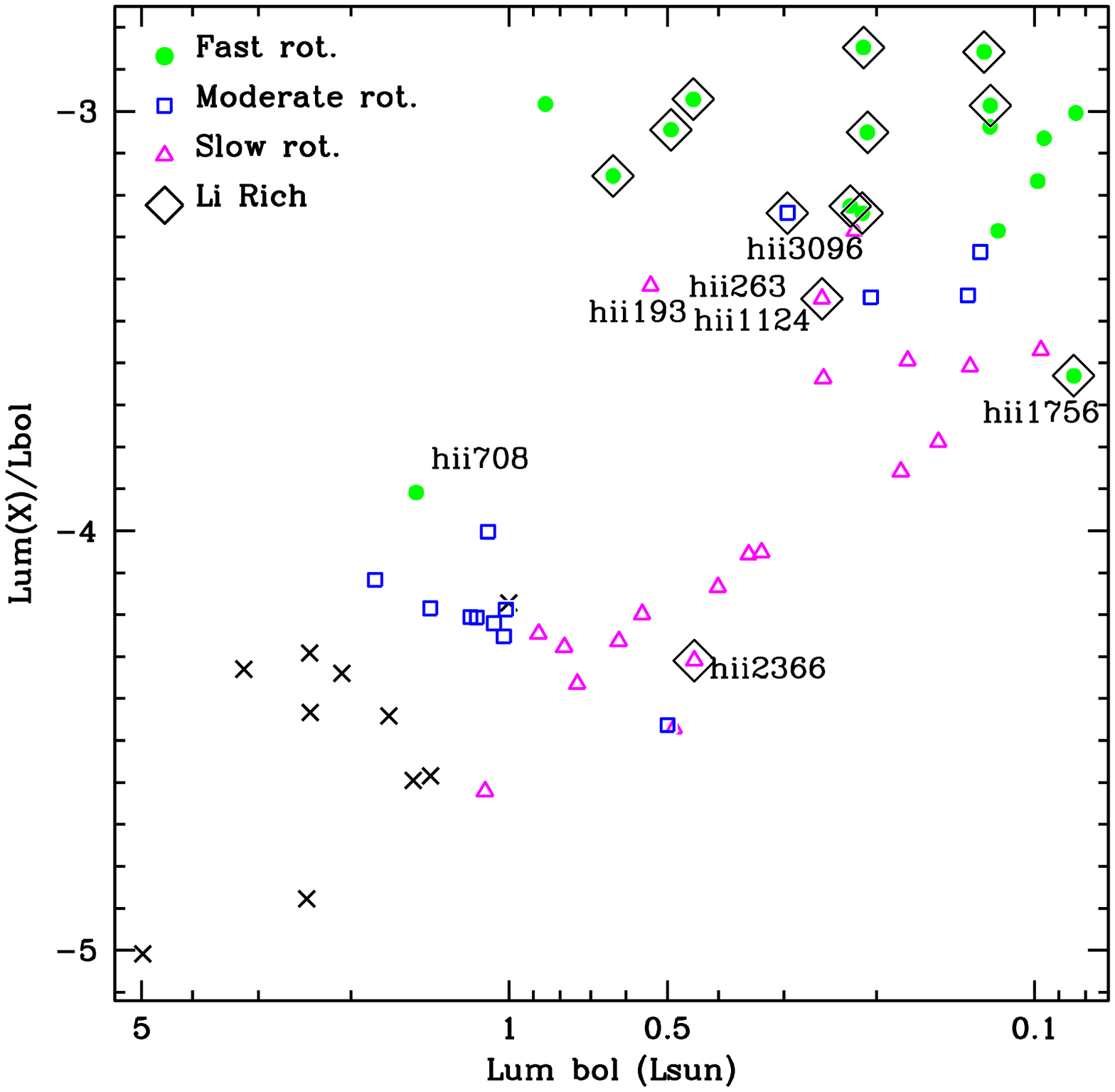} 
\includegraphics[width=0.5\textwidth,scale=0.50]{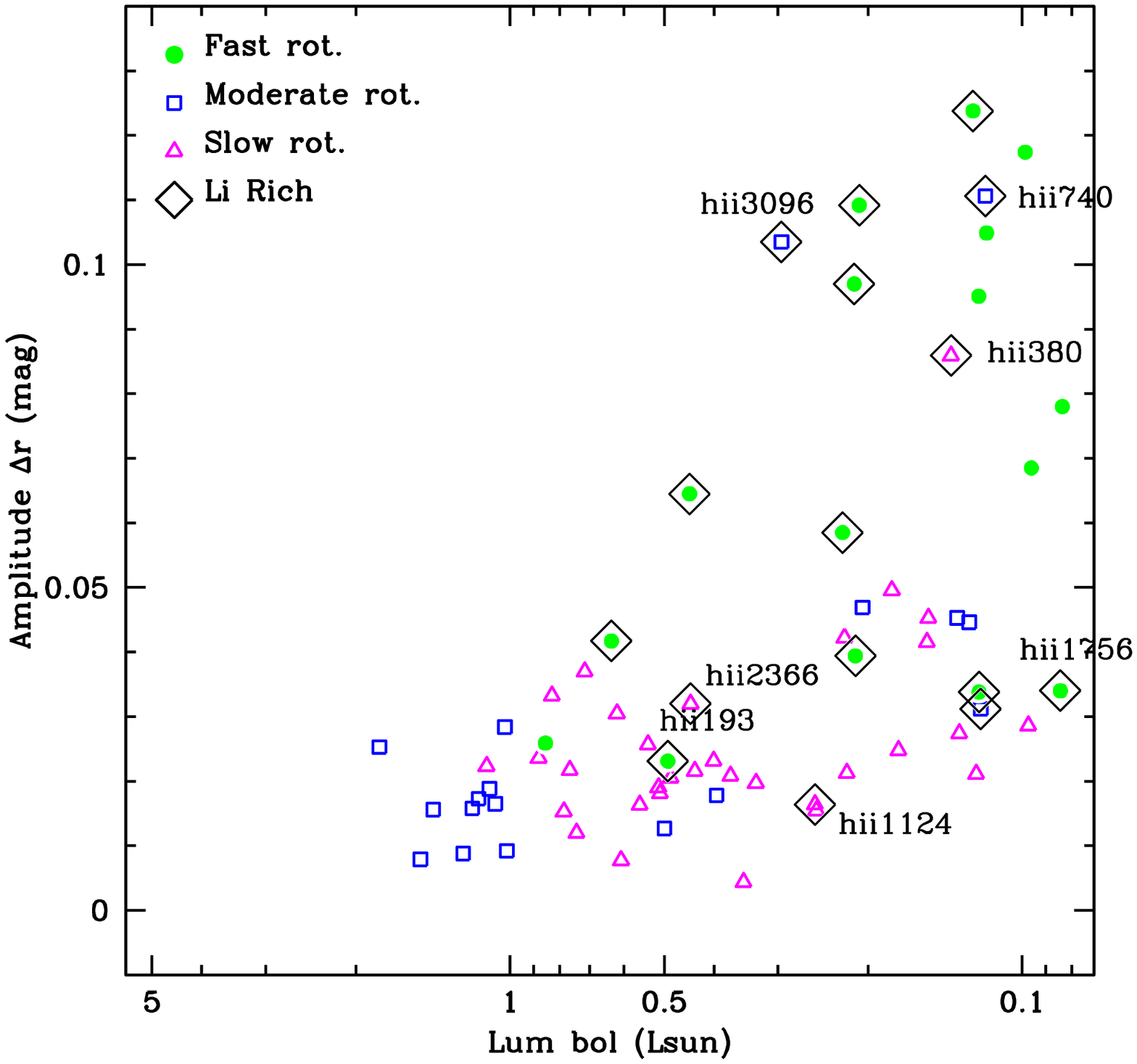} 
\caption{\label{LumBol_LxLbol_and_Ampl} 
%, 
{\bf Top.-} Comparison of the X-ray and the bolometric luminosities, taking into account the rotational periods.
 Note that HII380 and HII740 do not have a measured value of Lx and therefore they are not
 represented on the diagram.
{\bf Bottom.-} The amplitude of the light curve.
Only single stars and the W(Li) based on the spectral resolution are displayed in both panels. 
}
\end{figure}

In any event, just to verify whether the stellar mass might play a hidden role in the trends discussed above, 
Fig. \ref{LumBol_LxLbol_and_Ampl}a compares the ratio Lum $L_{X}$/$L_{Bol}$ with the bolometric luminosity (as a proxy of the stellar mass).
Clearly, the lithium excesses are related with X-ray overluminosities: the connection rotation-activity-lithium is striking, specially for the luminosity range
0.1-1 $L_\odot$ (up to 1.1 $M_\odot$ or G5 spectral type).
Note, however, that the X-ray is coming from the corona  and the lithium feature is formed at the photosphere.
 Thus, the optical variability should be a better indicator of the effect of activity on the observed lithium equivalent width.
Therefore, as a cross-check, we have used the  amplitude of the light-curve and compared with the bolometric luminosity  
(Fig. \ref{LumBol_LxLbol_and_Ampl}b), which support our interpretation.

 Thus, although there is no one-to-one relation between lithium over-abundance and rotation and/or activity, we conclude that part of the origin of the observed lithium spread could be  activity via surface inhomogeneities. 
 This effect is partially hidden by other factors such as
spot filling factor, the inclination of the rotation, and binarity,
 but the trend is significant.
 Note, however, that a much younger association, namely NGC2264
 displays a clear connection between rotation and lithium which should be intrinsic and cannot be connected to activity,
 since this 5 Myr sample is saturated  (see details in \citealp{Bouvier2016-Li-Rotation-NGC2264}).
 Thus, both factors should be playing a role
 although the relative importance of these two effects could change at different ages.

%%%%%%%%%%%%%%%%%%%%%%%%%%%%%%%%%%%%%%%%%%%%%%%%%%%%%%%%%%%%%%%%%%%%%%%%%%%%%%%%%%%%%%%%%%%%%
%
\subsection{The lithium-chronology: lithium depletion pattern for slow and fast rotators \label{subsub:LiPattern_slowFast}}
%
%%%%%%%%%%%%%%%%%%%%%%%%%%%%%%%%%%%%%%%%%%%%%%%%%%%%%%%%%%%%%%%%%%%%%%%%%%%%%%%%%%%%%%%%%%%%%

In order to derive an age estimate for  single stars or a new moving group,
it is quite common to compare their lithium equivalent widths with the
lithium depletion pattern for several well known open clusters and stellar associations at different evolutionary states.
However, for a given mass/$T_{\rm eff}$/luminosity, the visual inspection indicates that there is a large spread in W(Li),
 and there are not clear boundaries between clusters of very different ages. 

Since we have clearly established the relation between W(Li) and rotation, independently of the original reason, we propose
to make this type of comparison (the lithium-chronology for G and K stars) taking into account the rotation class. A slow rotator
should be compared solely with cluster members with large rotational periods and the same holds for fast rotators.
Age estimates derived this way should have much smaller error bars.

%%%%%%%%%%%%%%%%%%%%%%%%%%%%%%%%%%%%%%%%%%%%%%%%%%%%%%%%%%%%%%%%%%%%%%%%%%%%%%%%%%%%%%%%%%%%%

%%%%%%%%%%%%%%%%%%%%%%%%%%%%%%%%%%%%%%%%%%%%%%%%%%%%%%%%%%%%%%%%%%%%%%%%%%%%%%%%%%%%%%%%%%%%%
%
\subsection{Pre-Main Sequence evolution: lithium, rotation, radii and planetary systems\label{subsub:PMS_LiPlanets}}
%
%%%%%%%%%%%%%%%%%%%%%%%%%%%%%%%%%%%%%%%%%%%%%%%%%%%%%%%%%%%%%%%%%%%%%%%%%%%%%%%%%%%%%%%%%%%%%

There is a  trend between planet host and  lithium depletion for solar-like stars (\citealp{Israelian2009_EnhancedLi}),
in the sense that stars with known planets are, on average, lithium-poor compared with similar stars.
 Note, however, that it has not been proved that the relation is physical and much work has to be done.  If real, the origin of this trend is not understood.
On the other hand, there are evidences of planets being engulfed or very close to it 
(\citealp{LilloBox2014.Kepler91_REB}, \citealp{LilloBox2014.Kepler91_RV})
by their host star, leaving abundance abnormalities (\citealp{Israelian2001_PlaentEngulfment}), but they would be enriched by metallic elements. 
If the trend is real, our interpretation is not related with contamination, but with the early evolution and the connection 
between the circumstellar disk and rotation (and activity).
Weak-line TTauri stars (i.e., diskless) rotate faster than Classical TTauri stars (with circumstellar disk), due to the disk-locking
produced by  magnetic fields. The dichotomy for the rotation rates are kept as stars evolved 
towards the Main Sequence (\citealp{Bouvier1997.1}). Once there, the slow rotators have a tendency to have debris disk, 
the assumed remnant of planet formation, suggesting that they do have a planetary system
 (\citealp{Bouvier2008.1}; \citealp{McQuillan2013.1}, 
\citealp{McQuillan2013.2}).
 As we have seen, Pleiades members, already very near the ZAMS, display a clear correlation between rotation and lithium equivalent width.
 The cause can be either and early effect, intrinsic effect due to rotation (as in the case of NGC2264, \citealp{Bouvier2016-Li-Rotation-NGC2264});
 and inflated radii due to fast rotation producing a reduced lithium depletion
 (as postulated by \citealp{Somers2014-LithiumPMS}, implying a different evolutionary history);
 or an apparent effect due to the rotation-activity connection, a cooler effective temperature and enhanced
lithium equivalent width, without modifying the lithium abundance depletion pattern.
Independently of the validity of these three scenarios (perhaps all are at work), 
the distribution of rotational periods and lithium in Pleiades bona-fide members suggest that
the likelihood of finding planetary systems in slow rotators, with lower W(Li), may be much larger than in the
case of lithium-rich, fast rotators. Moreover, since the $vsini$ and the activity are small, 
from the technical point of view it would be easier to find them.

%%%%%%%%%%%%%%%%%%%%%%%%%%%%%%%%%%%%%%%%%%%%%%%%%%%%%%%%%%%%%%%%%%%%%%%%%%%%%%%%%%%%%%%%%%%%%  SECTION
%%%%%%%%%%%%%%%%%%%%%%%%%%%%%%%%%%%%%%%%%%%%%%%%%%%%%%%%%%%%%%%%%%%%%%%%%%%%%%%%%%%%%%%%%%%%%
%%%%%%%%%%%%%%%%%%%%%%%%%%%%%%%%%%%%%%%%%%%%%%%%%%%%%%%%%%%%%%%%%%%%%%%%%%%%%%%%%%%%%%%%%%%%%
%
\section{Conclusions\label{sec:conclusions}}
%
%%%%%%%%%%%%%%%%%%%%%%%%%%%%%%%%%%%%%%%%%%%%%%%%%%%%%%%%%%%%%%%%%%%%%%%%%%%%%%%%%%%%%%%%%%%%%
%%%%%%%%%%%%%%%%%%%%%%%%%%%%%%%%%%%%%%%%%%%%%%%%%%%%%%%%%%%%%%%%%%%%%%%%%%%%%%%%%%%%%%%%%%%%%
%%%%%%%%%%%%%%%%%%%%%%%%%%%%%%%%%%%%%%%%%%%%%%%%%%%%%%%%%%%%%%%%%%%%%%%%%%%%%%%%%%%%%%%%%%%%%
%%%%%%%%%%%%%%%%%%%%%%%%%%%%

We have compiled all the available lithium equivalent with in members of the Pleiades open cluster and reassessed the membership
probability based on the DANCe database 
(\citealp{Sarro2014.1}; 
\citealp{Bouy2015-Pleiades}).  
In addition, we have compiled multi-wavelength photometry from DANCe and from 
the literature based on the Virtual Observatory using capabilities of VOSA (\citealp{Bayo2008-VOSA}),
in order to derive accurate bolometric luminosities and effective temperatures.
In addition, we  have gathered all the available information regarding rotational periods, projected equatorial velocities, binarity and activity.
Thus, we selected a sub-sample of Pleiades members based on their position on the Herzprung-Russell diagram
and the membership probability as derived from proper motions and different photometric Color-Magnitude Diagrams.

We have not found any undisputed trend between a star location in the HR diagram and the rotation. If there is, it might be connected
with activity and the presence of large coverage by stellar spots, which should affect the derived effective temperature and/or 
the bolometric luminosity. However, the relation between lithium equivalent with and rotation is much more evident, specially for single
stars with the same luminosity in the range 
 $Lum(bol)=0.5-0.2$ $L_\odot$ (about K2-K7 spectral types  or 0.75--0.95 $M_\sun$), but even reaching
1.0 $L_\odot$ (about G5 or 1.1 $M_\sun$).
This lithium excess,
 which is connected to rotation (\citealp{Butler1987-Rotation-Pleiades}; \citealp{Soderblom1993-Li-Pleiades}), 
also seems to be connected with activity  (initially suggested by \citealp{GarciaLopez1991.2}),
either in terms of photospheric variability (the amplitude) 
and X-ray emission (coming from the stellar corona).
Thus, although we cannot find a bijective relation either between lithium excess and rotation or activity
(in the sense that star with lithium excess are {\it always} very active and rotate fast and {\it vice versa}), 
it seems that part of the explanation
falls on the last factor: surface inhomogeneities would modify the observed lithium equivalent. 
 Moreover, since a {\it bona fide} lithium enhancement is present in young, fast rotators (as the case of NGC2264 illustrates),
both activity and rotation should play a role in the lithium problem.
In addition, we suggest that, due to the bi- or multi-modal distribution of the rotational periods for a given mass, 
the gyro-chronology should take this fact into account.

Finally, it is possible to speculate that, {\bf if} there is a  connection between lithium and the presence of
 planets in field stars, Pleiades members with low lithium and slow rotation might have a larger probability of harboring planetary systems.

  In any case, since the lithium surface abundance  is a probe to the stellar interior and despite quite sophisticated recent theoretical  efforts we do not understand its evolution, it seems necessary to revisit the problem with new, homogeneous data taken with a large spectral resolution instrument at a very high signal to noise ratio, coupled with a complete data-set (rotation, activity, other alkali and so on). Thus, only by having all this information at hand we can expected to be able to provide a comprehensive interpretation  and the problem be solved.

\begin{acknowledgements}
 This research has been funded by Spanish grant ESP2015-65712-C5-1-R,  as well as by the French
grant ANR-10-JCJC-0501 DESC, as well as by the CNRS PICS project
``Comprendre l'origine et les proprietes des etoiles jeunes''.
HB is funded by the Spanish Ram\'on y Cajal fellowship program number RYC-2009-04497, whereas
JB and EM acknowledge financial support from the grant ANR 2011 Blanc SIMI5-6 020 01 “Toupies: Towards understanding the spin evolution of stars”.
We also  acknowledge support from the Faculty of the European Space Astronomy Centre (ESAC).
It makes use of VOSA, developed under the Spanish Virtual Observatory project
 supported from the Spanish MICINN through grant AyA2008-02156, and
 of the SIMBAD database, operated at CDS, Strasbourg, France.
This work is based in part on data obtained as part of the UKIRT Infrared Deep Sky Survey; 
 from the Wide-Field Infrared Survey Explorer, which is a joint project of the University of
California, Los Angeles, and the Jet Propulsion Laboratory/California
Institute of Technology, funded by the National Aeronautics and
Space Administration;
and data from the Two Micron All
Sky Survey, which is a joint project of the University of Massachusetts
and the Infrared Processing and Analysis Center/California Institute
of Technology, funded by the National Aeronautics and Space
Administration and the National Science Foundation.
We do greatly appreciate the comments and suggestions bt the anonymous referee.
\end{acknowledgements}

%%%%%%%%%%%%%%%%%%%%%%%%%%%%%%%%%%%%%%%%%%%%%%%%%%%%%%%%%%%%%%
%%%%%%%%%%%%%%%%%%%%%%%%%%%%%%%%%%%%%%%%%%%%%%%%%%%%%%%%%%%%%%
%%%%%%%%%%%%%%%%%%%%%%%%%%%%%%%%%%%%%%%%%%%%%%%%%%%%%%%%%%%%%%
%%%%%%%%%%%%%%%%%%%%%%%%%%%%%%%%%%%%%%%%%%%%%%%%%%%%%%%%%%%%%%
%%%%%%%%%%%%%%%%%%%%%%%%%%%%%%%%%%%%%%%%%%%%%%%%%%%%%%%%%%%%%%
\bibliographystyle{aa} %aa.bst
%\bibliography{0_LDB}
\bibliography{00_bibliography}

\begin{thebibliography}{119}
\expandafter\ifx\csname natexlab\endcsname\relax\def\natexlab#1{#1}\fi

\bibitem[{{Adelman-McCarthy} \& {et al.}(2011)}]{SDSS2011}
{Adelman-McCarthy}, J.~K. \& {et al.} 2011, VizieR Online Data Catalog, 2306, 0

\bibitem[{{Allard} {et~al.}(2012){Allard}, {Homeier}, {Freytag}, \&
  {Sharp}}]{Allard2012.1}
{Allard}, F., {Homeier}, D., {Freytag}, B., \& {Sharp}, C.~M. 2012, in EAS
  Publications Series, Vol.~57, EAS Publications Series, ed. C.~{Reyl{\'e}},
  C.~{Charbonnel}, \& M.~{Schultheis}, 3--43

\bibitem[{{Alves} \& {Bouy}(2012)}]{Alves2012.1}
{Alves}, J. \& {Bouy}, H. 2012, \aap, 547, A97

\bibitem[{{Argiroffi} {et~al.}(2013){Argiroffi}, {Caramazza}, {Micela},
  {Moraux}, \& {Bouvier}}]{Argiroffi2013.1}
{Argiroffi}, C., {Caramazza}, M., {Micela}, G., {Moraux}, E., \& {Bouvier}, J.
  2013, Protostars and Planets VI, 89

\bibitem[{{Balachandran}(1995)}]{Balachandran1995-Li-F-M67}
{Balachandran}, S. 1995, \apj, 446, 203

\bibitem[{{Baraffe} \& {Chabrier}(2010)}]{Baraffe2010.1}
{Baraffe}, I. \& {Chabrier}, G. 2010, \aap, 521, A44

\bibitem[{{Barrado y Navascu{\'e}s} {et~al.}(2001{\natexlab{a}}){Barrado y
  Navascu{\'e}s}, {Deliyannis}, \& {Stauffer}}]{Barrado2001-Li-M35}
{Barrado y Navascu{\'e}s}, D., {Deliyannis}, C.~P., \& {Stauffer}, J.~R.
  2001{\natexlab{a}}, \apj, 549, 452

\bibitem[{{Barrado y Navascues} {et~al.}(1997){Barrado y Navascues},
  {Fernandez-Figueroa}, {Garcia Lopez}, {de Castro}, \&
  {Cornide}}]{Barrado1997-Age-Dwarfs-RSCVn}
{Barrado y Navascues}, D., {Fernandez-Figueroa}, M.~J., {Garcia Lopez}, R.~J.,
  {de Castro}, E., \& {Cornide}, M. 1997, \aap, 326, 780

\bibitem[{{Barrado y Navascu{\'e}s} {et~al.}(2001{\natexlab{b}}){Barrado y
  Navascu{\'e}s}, {Garc{\'{\i}}a L{\'o}pez}, {Severino}, \&
  {Gomez}}]{Barrado2001-Li-Na-Ka-Activity-Pleiades}
{Barrado y Navascu{\'e}s}, D., {Garc{\'{\i}}a L{\'o}pez}, R.~J., {Severino},
  G., \& {Gomez}, M.~T. 2001{\natexlab{b}}, \aap, 371, 652

\bibitem[{{Barrado y Navascues} \& {Stauffer}(1996)}]{Barrado1996-Li-Hyades}
{Barrado y Navascues}, D. \& {Stauffer}, J.~R. 1996, \aap, 310, 879

\bibitem[{{Barrado y Navascu{\'e}s} {et~al.}(2004){Barrado y Navascu{\'e}s},
  {Stauffer}, \& {Jayawardhana}}]{Barrado2004-Li-Ha-IC2391}
{Barrado y Navascu{\'e}s}, D., {Stauffer}, J.~R., \& {Jayawardhana}, R. 2004,
  \apj, 614, 386

\bibitem[{{Barrado y Navascu{\'e}s} {et~al.}(1999){Barrado y Navascu{\'e}s},
  {Stauffer}, \& {Patten}}]{Barrado1999-Li-LDB-IC2391}
{Barrado y Navascu{\'e}s}, D., {Stauffer}, J.~R., \& {Patten}, B.~M. 1999,
  \apjl, 522, L53

\bibitem[{{Basri} {et~al.}(1996){Basri}, {Marcy}, \& {Graham}}]{Basri1996.1}
{Basri}, G., {Marcy}, G.~W., \& {Graham}, J.~R. 1996, \apj, 458, 600

\bibitem[{{Bayo} {et~al.}(2012){Bayo}, {Barrado}, {Hu{\'e}lamo},
  {Morales-Calder{\'o}n}, {Melo}, {Stauffer}, \&
  {Stelzer}}]{Bayo2012-Li-Rotation-Activity-C69}
{Bayo}, A., {Barrado}, D., {Hu{\'e}lamo}, N., {et~al.} 2012, \aap, 547, A80

\bibitem[{{Bayo} {et~al.}(2008){Bayo}, {Rodrigo}, {Barrado y Navascu{\'e}s},
  {Solano}, {Guti{\'e}rrez}, {Morales-Calder{\'o}n}, \&
  {Allard}}]{Bayo2008-VOSA}
{Bayo}, A., {Rodrigo}, C., {Barrado y Navascu{\'e}s}, D., {et~al.} 2008, \aap,
  492, 277

\bibitem[{{Bianchi} {et~al.}(2011){Bianchi}, {Herald}, {Efremova}, {Girardi},
  {Zabot}, {Marigo}, {Conti}, \& {Shiao}}]{Bianchi2011.1}
{Bianchi}, L., {Herald}, J., {Efremova}, B., {et~al.} 2011, \apss, 335, 161

\bibitem[{{Boesgaard}(1987{\natexlab{a}})}]{Boesgaard1987-Li-F-Hyades}
{Boesgaard}, A.~M. 1987{\natexlab{a}}, \pasp, 99, 1067

\bibitem[{{Boesgaard}(1987{\natexlab{b}})}]{Boesgaard1987-Li-Coma}
{Boesgaard}, A.~M. 1987{\natexlab{b}}, \apj, 321, 967

\bibitem[{{Boesgaard} \& {Budge}(1988)}]{Boesgaard1988-Li-HyadesMG-Praesepe}
{Boesgaard}, A.~M. \& {Budge}, K.~G. 1988, \apj, 332, 410

\bibitem[{{Boesgaard} \& {Budge}(1989)}]{Boesgaard1989-LiBe-Hyades}
{Boesgaard}, A.~M. \& {Budge}, K.~G. 1989, \apj, 338, 875

\bibitem[{{Boesgaard} {et~al.}(1988){Boesgaard}, {Budge}, \&
  {Ramsay}}]{Boesgaard1988-Li-Pleiade-APer}
{Boesgaard}, A.~M., {Budge}, K.~G., \& {Ramsay}, M.~E. 1988, \apj, 327, 389

\bibitem[{{Boesgaard} \&
  {Tripicco}(1986{\natexlab{a}})}]{Boesgaard1986-Li-Fdwarfs}
{Boesgaard}, A.~M. \& {Tripicco}, M.~J. 1986{\natexlab{a}}, \apj, 303, 724

\bibitem[{{Boesgaard} \&
  {Tripicco}(1986{\natexlab{b}})}]{Boesgaard1986-Li-Hyades}
{Boesgaard}, A.~M. \& {Tripicco}, M.~J. 1986{\natexlab{b}}, \apjl, 302, L49

\bibitem[{{Bouvier}(2008)}]{Bouvier2008.1}
{Bouvier}, J. 2008, \aap, 489, L53

\bibitem[{{Bouvier} {et~al.}(1997{\natexlab{a}}){Bouvier}, {Forestini}, \&
  {Allain}}]{Bouvier1997.1}
{Bouvier}, J., {Forestini}, M., \& {Allain}, S. 1997{\natexlab{a}}, \aap, 326,
  1023

\bibitem[{{Bouvier} {et~al.}(2016){Bouvier}, {Lanzafame}, {Venuti}, {Klutsch},
  {Jeffries}, {Frasca}, {Moraux}, {Biazzo}, {Messina}, {Micela}, {Randich},
  {Stauffer}, {Cody}, {Flaccomio}, {Gilmore}, {Bayo}, {Bensby}, {Bragaglia},
  {Carraro}, {Casey}, {Costado}, {Damiani}, {Delgado Mena}, {Donati},
  {Franciosini}, {Hourihane}, {Koposov}, {Lardo}, {Lewis}, {Magrini}, {Monaco},
  {Morbidelli}, {Prisinzano}, {Sacco}, {Sbordone}, {Sousa}, {Vallenari},
  {Worley}, {Zaggia}, \& {Zwitter}}]{Bouvier2016-Li-Rotation-NGC2264}
{Bouvier}, J., {Lanzafame}, A.~C., {Venuti}, L., {et~al.} 2016, \aap, 590, A78

\bibitem[{{Bouvier} {et~al.}(1997{\natexlab{b}}){Bouvier}, {Rigaut}, \&
  {Nadeau}}]{Bouvier1997.Pleiades_binaries}
{Bouvier}, J., {Rigaut}, F., \& {Nadeau}, D. 1997{\natexlab{b}}, \aap, 323, 139

\bibitem[{{Bouy} {et~al.}(2014){Bouy}, {Alves}, {Bertin}, {Sarro}, \&
  {Barrado}}]{Bouy2014-Orion}
{Bouy}, H., {Alves}, J., {Bertin}, E., {Sarro}, L.~M., \& {Barrado}, D. 2014,
  \aap, 564, A29

\bibitem[{{Bouy} {et~al.}(2013){Bouy}, {Bertin}, {Moraux}, {Cuillandre},
  {Bouvier}, {Barrado}, {Solano}, \& {Bayo}}]{Bouy2013-DANCE}
{Bouy}, H., {Bertin}, E., {Moraux}, E., {et~al.} 2013, \aap, 554, A101

\bibitem[{{Bouy} {et~al.}(2015){Bouy}, {Bertin}, {Sarro}, {Barrado}, {Moraux},
  {Bouvier}, {Cuillandre}, {Berihuete}, {Olivares}, \&
  {Beletsky}}]{Bouy2015-Pleiades}
{Bouy}, H., {Bertin}, E., {Sarro}, L.~M., {et~al.} 2015, \aap, 577, A148

\bibitem[{{Butler} {et~al.}(1987){Butler}, {Marcy}, {Cohen}, \&
  {Duncan}}]{Butler1987-Rotation-Pleiades}
{Butler}, R.~P., {Marcy}, G.~W., {Cohen}, R.~D., \& {Duncan}, D.~K. 1987,
  \apjl, 319, L19

\bibitem[{{Castelli} {et~al.}(1997){Castelli}, {Gratton}, \&
  {Kurucz}}]{castelli97}
{Castelli}, F., {Gratton}, R.~G., \& {Kurucz}, R.~L. 1997, \aap, 318, 841

\bibitem[{{Chabrier} {et~al.}(2000){Chabrier}, {Baraffe}, {Allard}, \&
  {Hauschildt}}]{Chabrier2000.1}
{Chabrier}, G., {Baraffe}, I., {Allard}, F., \& {Hauschildt}, P. 2000, \apj,
  542, 464

\bibitem[{{Cutri} \& {et al.}(2012)}]{Cutri2012.1}
{Cutri}, R.~M. \& {et al.} 2012, VizieR Online Data Catalog, 2311, 0

\bibitem[{{Cutri} {et~al.}(2003){Cutri}, {Skrutskie}, {van Dyk}, {Beichman},
  {Carpenter}, {Chester}, {Cambresy}, {Evans}, {Fowler}, {Gizis}, {Howard},
  {Huchra}, {Jarrett}, {Kopan}, {Kirkpatrick}, {Light}, {Marsh}, {McCallon},
  {Schneider}, {Stiening}, {Sykes}, {Weinberg}, {Wheaton}, {Wheelock}, \&
  {Zacarias}}]{Cutri2003.1}
{Cutri}, R.~M., {Skrutskie}, M.~F., {van Dyk}, S., {et~al.} 2003, {2MASS All
  Sky Catalog of point sources.} (The IRSA 2MASS All-Sky Point Source Catalog,
  NASA/IPAC Infrared Science
  Archive.~http://irsa.ipac.caltech.edu/applications/Gator/)

\bibitem[{{Dahm}(2015)}]{Dahm2015-Pleiades-LDB}
{Dahm}, S.~E. 2015, \apj, 813, 108

\bibitem[{{Dolan} \& {Mathieu}(1999)}]{Dolan1999-Li-C69}
{Dolan}, C.~J. \& {Mathieu}, R.~D. 1999, \aj, 118, 2409

\bibitem[{{Ducourant} {et~al.}(2005){Ducourant}, {Teixeira}, {P{\'e}ri{\'e}},
  {Lecampion}, {Guibert}, \& {Sartori}}]{Ducourant2005.1}
{Ducourant}, C., {Teixeira}, R., {P{\'e}ri{\'e}}, J.~P., {et~al.} 2005, \aap,
  438, 769

\bibitem[{{Eggen}(1975)}]{Eggen1975.1}
{Eggen}, O.~J. 1975, \pasp, 87, 37

\bibitem[{{Eggen}(1995)}]{Eggen1995.1}
{Eggen}, O.~J. 1995, \aj, 110, 1749

\bibitem[{{Eggenberger} {et~al.}(2012){Eggenberger}, {Haemmerl{\'e}}, {Meynet},
  \& {Maeder}}]{Eggenberger2012-Li-Rotation-Disk-PMS}
{Eggenberger}, P., {Haemmerl{\'e}}, L., {Meynet}, G., \& {Maeder}, A. 2012,
  \aap, 539, A70

\bibitem[{{Garc{\'{\i}}a L{\'o}pez} {et~al.}(1991){Garc{\'{\i}}a L{\'o}pez},
  {Rebolo}, {Beckman}, \& {Magazz{\`u}}}]{GarciaLopez1991.2}
{Garc{\'{\i}}a L{\'o}pez}, R.~J., {Rebolo}, R., {Beckman}, J.~E., \&
  {Magazz{\`u}}, A. 1991, in Lecture Notes in Physics, Berlin Springer Verlag,
  Vol. 380, IAU Colloq. 130: The Sun and Cool Stars. Activity, Magnetism,
  Dynamos, ed. I.~{Tuominen}, D.~{Moss}, \& G.~{R{\"u}diger}, 443

\bibitem[{{Garcia Lopez} {et~al.}(1994){Garcia Lopez}, {Rebolo}, \&
  {Martin}}]{GarciaLopez1994.1}
{Garcia Lopez}, R.~J., {Rebolo}, R., \& {Martin}, E.~L. 1994, \aap, 282, 518

\bibitem[{{Gilmore} {et~al.}(2012){Gilmore}, {Randich}, {Asplund}, {Binney},
  {Bonifacio}, {Drew}, {Feltzing}, {Ferguson}, {Jeffries}, {Micela},
  {Negueruela}, {Prusti}, {Rix}, {Vallenari}, {Alfaro}, {Allende-Prieto},
  {Babusiaux}, {Bensby}, {Blomme}, {Bragaglia}, {Flaccomio}, {Fran{\c c}ois},
  {Irwin}, {Koposov}, {Korn}, {Lanzafame}, {Pancino}, {Paunzen},
  {Recio-Blanco}, {Sacco}, {Smiljanic}, {Van Eck}, \&
  {Walton}}]{Gilmore2012.GaiaESO}
{Gilmore}, G., {Randich}, S., {Asplund}, M., {et~al.} 2012, The Messenger, 147,
  25

\bibitem[{{Hambly} {et~al.}(1993){Hambly}, {Hawkins}, \&
  {Jameson}}]{Hambly1993.1}
{Hambly}, N.~C., {Hawkins}, M.~R.~S., \& {Jameson}, R.~F. 1993, \aaps, 100, 607

\bibitem[{{Haro} {et~al.}(1982){Haro}, {Chavira}, \& {Gonzalez}}]{Haro1982.1}
{Haro}, G., {Chavira}, E., \& {Gonzalez}, G. 1982, Boletin del Instituto de
  Tonantzintla, 3, 3

\bibitem[{{Hartman} {et~al.}(2010){Hartman}, {Bakos}, {Kov{\'a}cs}, \&
  {Noyes}}]{Hartman2010.1}
{Hartman}, J.~D., {Bakos}, G.~{\'A}., {Kov{\'a}cs}, G., \& {Noyes}, R.~W. 2010,
  \mnras, 408, 475

\bibitem[{{Herbig}(1965)}]{Herbig1965-Li-FG}
{Herbig}, G.~H. 1965, \apj, 141, 588

\bibitem[{{Hertzsprung}(1947)}]{Hertzsprung1947.1}
{Hertzsprung}, E. 1947, Annalen van de Sterrewacht te Leiden, 19, A1

\bibitem[{{H{\o}g} {et~al.}(2000){H{\o}g}, {Fabricius}, {Makarov}, {Urban},
  {Corbin}, {Wycoff}, {Bastian}, {Schwekendiek}, \& {Wicenec}}]{Hog2000.1}
{H{\o}g}, E., {Fabricius}, C., {Makarov}, V.~V., {et~al.} 2000, \aap, 355, L27

\bibitem[{{Howell} {et~al.}(2014){Howell}, {Sobeck}, {Haas}, {Still},
  {Barclay}, {Mullally}, {Troeltzsch}, {Aigrain}, {Bryson}, {Caldwell},
  {Chaplin}, {Cochran}, {Huber}, {Marcy}, {Miglio}, {Najita}, {Smith},
  {Twicken}, \& {Fortney}}]{Howell2014-Kepler-K2}
{Howell}, S.~B., {Sobeck}, C., {Haas}, M., {et~al.} 2014, \pasp, 126, 398

\bibitem[{{H{\"u}nsch} {et~al.}(2004){H{\"u}nsch}, {Randich}, {Hempel},
  {Weidner}, \& {Schmitt}}]{Hunsch2004.1}
{H{\"u}nsch}, M., {Randich}, S., {Hempel}, M., {Weidner}, C., \& {Schmitt},
  J.~H.~M.~M. 2004, \aap, 418, 539

\bibitem[{{Israelian} {et~al.}(2009){Israelian}, {Delgado Mena}, {Santos},
  {Sousa}, {Mayor}, {Udry}, {Dom{\'{\i}}nguez Cerde{\~n}a}, {Rebolo}, \&
  {Randich}}]{Israelian2009_EnhancedLi}
{Israelian}, G., {Delgado Mena}, E., {Santos}, N.~C., {et~al.} 2009, \nat, 462,
  189

\bibitem[{{Israelian} {et~al.}(2001){Israelian}, {Santos}, {Mayor}, \&
  {Rebolo}}]{Israelian2001_PlaentEngulfment}
{Israelian}, G., {Santos}, N.~C., {Mayor}, M., \& {Rebolo}, R. 2001, \nat, 411,
  163

\bibitem[{{Jackson} \& {Jeffries}(2013)}]{Jackson2013-Spots-Size}
{Jackson}, R.~J. \& {Jeffries}, R.~D. 2013, \mnras, 431, 1883

\bibitem[{{Jackson} \& {Jeffries}(2014)}]{Jackson2014-Spots-Radii-PMS}
{Jackson}, R.~J. \& {Jeffries}, R.~D. 2014, \mnras, 441, 2111

\bibitem[{{Jeffries}(1999)}]{Jeffries1999-Li-Pleiades}
{Jeffries}, R.~D. 1999, \mnras, 309, 189

\bibitem[{{Jeffries} {et~al.}(2009){Jeffries}, {Jackson}, {James}, \&
  {Cargile}}]{Jeffries2009-Li-IC4665}
{Jeffries}, R.~D., {Jackson}, R.~J., {James}, D.~J., \& {Cargile}, P.~A. 2009,
  \mnras, 400, 317

\bibitem[{{Jeffries} {et~al.}(1998){Jeffries}, {James}, \&
  {Thurston}}]{Jeffries1998-LiRotation-NGC2516}
{Jeffries}, R.~D., {James}, D.~J., \& {Thurston}, M.~R. 1998, \mnras, 300, 550

\bibitem[{{Jeffries} {et~al.}(2003){Jeffries}, {Oliveira}, {Barrado y
  Navascu{\'e}s}, \& {Stauffer}}]{Jeffries2003-Li-NGC2547}
{Jeffries}, R.~D., {Oliveira}, J.~M., {Barrado y Navascu{\'e}s}, D., \&
  {Stauffer}, J.~R. 2003, \mnras, 343, 1271

\bibitem[{{Jones} {et~al.}(1996){Jones}, {Fischer}, \&
  {Stauffer}}]{Jones1996.1}
{Jones}, B.~F., {Fischer}, D.~A., \& {Stauffer}, J.~R. 1996, \aj, 112, 1562

\bibitem[{{King} \& {Hiltgen}(1996)}]{King1996-Li-Praesepe-TLBS}
{King}, J.~R. \& {Hiltgen}, D.~R. 1996, \pasp, 108, 246

\bibitem[{{King} {et~al.}(2000){King}, {Krishnamurthi}, \&
  {Pinsonneault}}]{King2000-Lithium-Rotation}
{King}, J.~R., {Krishnamurthi}, A., \& {Pinsonneault}, M.~H. 2000, \aj, 119,
  859

\bibitem[{{King} \& {Schuler}(2004)}]{King2004-Activity-Alkali}
{King}, J.~R. \& {Schuler}, S.~C. 2004, \aj, 128, 2898

\bibitem[{{King} {et~al.}(2010){King}, {Schuler}, {Hobbs}, \&
  {Pinsonneault}}]{King2010-Li-K-Scatter-Pleiades}
{King}, J.~R., {Schuler}, S.~C., {Hobbs}, L.~M., \& {Pinsonneault}, M.~H. 2010,
  \apj, 710, 1610

\bibitem[{{Lillo-Box} {et~al.}(2014{\natexlab{a}}){Lillo-Box}, {Barrado},
  {Henning}, {Mancini}, {Ciceri}, {Figueira}, {Santos}, {Aceituno}, \&
  {S{\'a}nchez}}]{LilloBox2014.Kepler91_RV}
{Lillo-Box}, J., {Barrado}, D., {Henning}, T., {et~al.} 2014{\natexlab{a}},
  \aap, 568, L1

\bibitem[{{Lillo-Box} {et~al.}(2014{\natexlab{b}}){Lillo-Box}, {Barrado},
  {Moya}, {Montesinos}, {Montalb{\'a}n}, {Bayo}, {Barbieri}, {R{\'e}gulo},
  {Mancini}, {Bouy}, \& {Henning}}]{LilloBox2014.Kepler91_REB}
{Lillo-Box}, J., {Barrado}, D., {Moya}, A., {et~al.} 2014{\natexlab{b}}, \aap,
  562, A109

\bibitem[{{Lodieu} {et~al.}(2012){Lodieu}, {Deacon}, \&
  {Hambly}}]{Lodieu2012.Pleiades}
{Lodieu}, N., {Deacon}, N.~R., \& {Hambly}, N.~C. 2012, \mnras, 422, 1495

\bibitem[{{Marcy} {et~al.}(1994){Marcy}, {Basri}, \& {Graham}}]{Marcy1994.1}
{Marcy}, G.~W., {Basri}, G., \& {Graham}, J.~R. 1994, \apjl, 428, L57

\bibitem[{{Margheim}(2007)}]{Margheim2007.1}
{Margheim}, S.~J. 2007, PhD thesis, Indiana University

\bibitem[{{Martin} {et~al.}(1998){Martin}, {Basri}, {Gallegos}, {Rebolo},
  {Zapatero-Osorio}, \& {Bejar}}]{Martin1998.1}
{Martin}, E.~L., {Basri}, G., {Gallegos}, J.~E., {et~al.} 1998, \apjl, 499, L61

\bibitem[{{Mart{\'{\i}}n} {et~al.}(2000){Mart{\'{\i}}n}, {Brandner}, {Bouvier},
  {Luhman}, {Stauffer}, {Basri}, {Zapatero Osorio}, \& {Barrado y
  Navascu{\'e}s}}]{Martin2000.1}
{Mart{\'{\i}}n}, E.~L., {Brandner}, W., {Bouvier}, J., {et~al.} 2000, \apj,
  543, 299

\bibitem[{{Mart{\'{\i}}n} {et~al.}(2005){Mart{\'{\i}}n}, {Magazz{\`u}},
  {Garc{\'{\i}}a L{\'o}pez}, {Randich}, \& {Barrado y
  Navascu{\'e}s}}]{Martin2005.1}
{Mart{\'{\i}}n}, E.~L., {Magazz{\`u}}, A., {Garc{\'{\i}}a L{\'o}pez}, R.~J.,
  {Randich}, S., \& {Barrado y Navascu{\'e}s}, D. 2005, \aap, 429, 1051

\bibitem[{{McQuillan} {et~al.}(2013{\natexlab{a}}){McQuillan}, {Aigrain}, \&
  {Mazeh}}]{McQuillan2013.2}
{McQuillan}, A., {Aigrain}, S., \& {Mazeh}, T. 2013{\natexlab{a}}, \mnras, 432,
  1203

\bibitem[{{McQuillan} {et~al.}(2013{\natexlab{b}}){McQuillan}, {Mazeh}, \&
  {Aigrain}}]{McQuillan2013.1}
{McQuillan}, A., {Mazeh}, T., \& {Aigrain}, S. 2013{\natexlab{b}}, \apjl, 775,
  L11

\bibitem[{{Melis} {et~al.}(2014){Melis}, {Reid}, {Mioduszewski}, {Stauffer}, \&
  {Bower}}]{Melis2014.1}
{Melis}, C., {Reid}, M.~J., {Mioduszewski}, A.~J., {Stauffer}, J.~R., \&
  {Bower}, G.~C. 2014, Science, 345, 1029

\bibitem[{{Mermilliod} {et~al.}(1992){Mermilliod}, {Rosvick}, {Duquennoy}, \&
  {Mayor}}]{Mermilliod1992_PleiadesBinarity}
{Mermilliod}, J.-C., {Rosvick}, J.~M., {Duquennoy}, A., \& {Mayor}, M. 1992,
  \aap, 265, 513

\bibitem[{{Michaud} \& {Charbonneau}(1991)}]{Michaud1991-Li}
{Michaud}, G. \& {Charbonneau}, P. 1991, \ssr, 57, 1

\bibitem[{{Oppenheimer} {et~al.}(1997){Oppenheimer}, {Basri}, {Nakajima}, \&
  {Kulkarni}}]{Oppenheimer1997.1}
{Oppenheimer}, B.~R., {Basri}, G., {Nakajima}, T., \& {Kulkarni}, S.~R. 1997,
  \aj, 113, 296

\bibitem[{{Palla} {et~al.}(2007){Palla}, {Randich}, {Pavlenko}, {Flaccomio}, \&
  {Pallavicini}}]{Palla2007.1}
{Palla}, F., {Randich}, S., {Pavlenko}, Y.~V., {Flaccomio}, E., \&
  {Pallavicini}, R. 2007, \apjl, 659, L41

\bibitem[{{Pasquini} {et~al.}(2008){Pasquini}, {Biazzo}, {Bonifacio},
  {Randich}, \& {Bedin}}]{Pasquini2008-M67-Li}
{Pasquini}, L., {Biazzo}, K., {Bonifacio}, P., {Randich}, S., \& {Bedin}, L.~R.
  2008, \aap, 489, 677

\bibitem[{{Pasquini} {et~al.}(1997){Pasquini}, {Randich}, \&
  {Pallavicini}}]{Pasquini1997-Lithium-M67}
{Pasquini}, L., {Randich}, S., \& {Pallavicini}, R. 1997, \aap, 325, 535

\bibitem[{{Pavlenko} \& {Magazzu}(1996)}]{Pavlenko1996.1}
{Pavlenko}, Y.~V. \& {Magazzu}, A. 1996, \aap, 311, 961

\bibitem[{{Perryman} {et~al.}(1997){Perryman}, {Lindegren}, {Kovalevsky},
  {Hoeg}, {Bastian}, {Bernacca}, {Cr{\'e}z{\'e}}, {Donati}, {Grenon}, {van
  Leeuwen}, {van der Marel}, {Mignard}, {Murray}, {Le Poole}, {Schrijver},
  {Turon}, {Arenou}, {Froeschl{\'e}}, \& {Petersen}}]{Perryman97}
{Perryman}, M.~A.~C., {Lindegren}, L., {Kovalevsky}, J., {et~al.} 1997, \aap,
  323, L49

\bibitem[{{Pilachowski}(1986)}]{Pilachowski1986-Li-NGC7789}
{Pilachowski}, C. 1986, \apj, 300, 289

\bibitem[{{Pilachowski} {et~al.}(1988){Pilachowski}, {Saha}, \&
  {Hobbs}}]{Pilachowski1988-L-NGC752-M67}
{Pilachowski}, C., {Saha}, A., \& {Hobbs}, L.~M. 1988, \pasp, 100, 474

\bibitem[{{Pilachowski} {et~al.}(1987){Pilachowski}, {Booth}, \&
  {Hobbs}}]{Pilachowski1987-Li-F-Pleiades}
{Pilachowski}, C.~A., {Booth}, J., \& {Hobbs}, L.~M. 1987, \pasp, 99, 1288

\bibitem[{{Pilachowski} {et~al.}(1984){Pilachowski}, {Mould}, \&
  {Siegel}}]{Pilachowski1984-Li-NGC7789}
{Pilachowski}, C.~A., {Mould}, J.~R., \& {Siegel}, M.~J. 1984, \apjl, 282, L17

\bibitem[{{Pinfield} {et~al.}(2003){Pinfield}, {Dobbie}, {Jameson}, {Steele},
  {Jones}, \& {Katsiyannis}}]{Pinfield2003.1}
{Pinfield}, D.~J., {Dobbie}, P.~D., {Jameson}, R.~F., {et~al.} 2003, \mnras,
  342, 1241

\bibitem[{{Pinfield} {et~al.}(2000){Pinfield}, {Hodgkin}, {Jameson},
  {Cossburn}, {Hambly}, \& {Devereux}}]{Pinfield2000.1}
{Pinfield}, D.~J., {Hodgkin}, S.~T., {Jameson}, R.~F., {et~al.} 2000, \mnras,
  313, 347

\bibitem[{{Pinsonneault} {et~al.}(1998){Pinsonneault}, {Stauffer}, {Soderblom},
  {King}, \& {Hanson}}]{Pinsonneault1998.1}
{Pinsonneault}, M.~H., {Stauffer}, J., {Soderblom}, D.~R., {King}, J.~R., \&
  {Hanson}, R.~B. 1998, \apj, 504, 170

\bibitem[{{Queloz} {et~al.}(1998){Queloz}, {Allain}, {Mermilliod}, {Bouvier},
  \& {Mayor}}]{Queloz1998_PleiadesVrot}
{Queloz}, D., {Allain}, S., {Mermilliod}, J.-C., {Bouvier}, J., \& {Mayor}, M.
  1998, \aap, 335, 183

\bibitem[{{Randich}(2001)}]{Randich2001-Li-K-IC2602}
{Randich}, S. 2001, \aap, 377, 512

\bibitem[{{Randich} {et~al.}(2013){Randich}, {Gilmore}, \& {Gaia-ESO
  Consortium}}]{Randich2013.GaiaESO}
{Randich}, S., {Gilmore}, G., \& {Gaia-ESO Consortium}. 2013, The Messenger,
  154, 47

\bibitem[{{Randich} {et~al.}(2001){Randich}, {Pallavicini}, {Meola},
  {Stauffer}, \& {Balachandran}}]{Randich2001_Li_Metal_IC2602_IC2391}
{Randich}, S., {Pallavicini}, R., {Meola}, G., {Stauffer}, J.~R., \&
  {Balachandran}, S.~C. 2001, \aap, 372, 862

\bibitem[{{Rebolo} {et~al.}(1996){Rebolo}, {Martin}, {Basri}, {Marcy}, \&
  {Zapatero-Osorio}}]{Rebolo1996.1}
{Rebolo}, R., {Martin}, E.~L., {Basri}, G., {Marcy}, G.~W., \&
  {Zapatero-Osorio}, M.~R. 1996, \apjl, 469, L53

\bibitem[{{Rosvick} {et~al.}(1992){Rosvick}, {Mermilliod}, \&
  {Mayor}}]{Rosvick1992_PleiadesRV}
{Rosvick}, J.~M., {Mermilliod}, J.-C., \& {Mayor}, M. 1992, \aap, 255, 130

\bibitem[{{Sacco} {et~al.}(2015){Sacco}, {Jeffries}, {Randich}, {Franciosini},
  {Jackson}, {Cottaar}, {Spina}, {Palla}, {Mapelli}, {Alfaro}, {Bonito},
  {Damiani}, {Frasca}, {Klutsch}, {Lanzafame}, {Bayo}, {Barrado},
  {Jim{\'e}nez-Esteban}, {Gilmore}, {Micela}, {Vallenari}, {Allende Prieto},
  {Flaccomio}, {Carraro}, {Costado}, {Jofr{\'e}}, {Lardo}, {Magrini},
  {Morbidelli}, {Prisinzano}, \& {Sbordone}}]{Sacco2015.VelaOB2}
{Sacco}, G.~G., {Jeffries}, R.~D., {Randich}, S., {et~al.} 2015, \aap, 574, L7

\bibitem[{{Sarro} {et~al.}(2014){Sarro}, {Bouy}, {Berihuete}, {Bertin},
  {Moraux}, {Bouvier}, {Cuillandre}, {Barrado}, \& {Solano}}]{Sarro2014.1}
{Sarro}, L.~M., {Bouy}, H., {Berihuete}, A., {et~al.} 2014, \aap, 563, A45

\bibitem[{{Sergison} {et~al.}(2013){Sergison}, {Mayne}, {Naylor}, {Jeffries},
  \& {Bell}}]{Sergison2013.1}
{Sergison}, D.~J., {Mayne}, N.~J., {Naylor}, T., {Jeffries}, R.~D., \& {Bell},
  C.~P.~M. 2013, \mnras, 434, 966

\bibitem[{{Siess} {et~al.}(2000){Siess}, {Dufour}, \&
  {Forestini}}]{Siess2000.1}
{Siess}, L., {Dufour}, E., \& {Forestini}, M. 2000, \aap, 358, 593

\bibitem[{{Skrutskie} {et~al.}(2006){Skrutskie}, {Cutri}, {Stiening},
  {Weinberg}, {Schneider}, {Carpenter}, {Beichman}, {Capps}, {Chester},
  {Elias}, {Huchra}, {Liebert}, {Lonsdale}, {Monet}, {Price}, {Seitzer},
  {Jarrett}, {Kirkpatrick}, {Gizis}, {Howard}, {Evans}, {Fowler}, {Fullmer},
  {Hurt}, {Light}, {Kopan}, {Marsh}, {McCallon}, {Tam}, {Van Dyk}, \&
  {Wheelock}}]{2MASS2006}
{Skrutskie}, M.~F., {Cutri}, R.~M., {Stiening}, R., {et~al.} 2006, \aj, 131,
  1163

\bibitem[{{Soderblom} {et~al.}(1993{\natexlab{a}}){Soderblom}, {Fedele},
  {Jones}, {Stauffer}, \& {Prosser}}]{Soderblom1993-Li-Praesepe}
{Soderblom}, D.~R., {Fedele}, S.~B., {Jones}, B.~F., {Stauffer}, J.~R., \&
  {Prosser}, C.~F. 1993{\natexlab{a}}, \aj, 106, 1080

\bibitem[{{Soderblom} {et~al.}(1993{\natexlab{b}}){Soderblom}, {Jones},
  {Balachandran}, {Stauffer}, {Duncan}, {Fedele}, \&
  {Hudon}}]{Soderblom1993-Li-Pleiades}
{Soderblom}, D.~R., {Jones}, B.~F., {Balachandran}, S., {et~al.}
  1993{\natexlab{b}}, \aj, 106, 1059

\bibitem[{{Soderblom} {et~al.}(2005){Soderblom}, {Nelan}, {Benedict},
  {McArthur}, {Ramirez}, {Spiesman}, \&
  {Jones}}]{Soderblom2005-PleiadesDistance}
{Soderblom}, D.~R., {Nelan}, E., {Benedict}, G.~F., {et~al.} 2005, \aj, 129,
  1616

\bibitem[{{Soderblom} {et~al.}(1993{\natexlab{c}}){Soderblom}, {Pilachowski},
  {Fedele}, \& {Jones}}]{Soderblom1993-Li-UMaG}
{Soderblom}, D.~R., {Pilachowski}, C.~A., {Fedele}, S.~B., \& {Jones}, B.~F.
  1993{\natexlab{c}}, \aj, 105, 2299

\bibitem[{{Soderblom} {et~al.}(1993{\natexlab{d}}){Soderblom}, {Stauffer},
  {Hudon}, \& {Jones}}]{Soderblom1993-RotationActivity-FGK-Pleiades}
{Soderblom}, D.~R., {Stauffer}, J.~R., {Hudon}, J.~D., \& {Jones}, B.~F.
  1993{\natexlab{d}}, \apjs, 85, 315

\bibitem[{{Somers} \& {Pinsonneault}(2014)}]{Somers2014-LithiumPMS}
{Somers}, G. \& {Pinsonneault}, M.~H. 2014, \apj, 790, 72

\bibitem[{{Somers} \& {Pinsonneault}(2015)}]{Somers2015-LithiumPleiades}
{Somers}, G. \& {Pinsonneault}, M.~H. 2015, \mnras, 449, 4131

\bibitem[{{Stauffer} \& {Hartmann}(1987)}]{Stauffer1987-rotation-Pleiades}
{Stauffer}, J.~R. \& {Hartmann}, L.~W. 1987, \apj, 318, 337

\bibitem[{{Stauffer} {et~al.}(2007){Stauffer}, {Hartmann}, {Fazio}, {Allen},
  {Patten}, {Lowrance}, {Hurt}, {Rebull}, {Cutri}, {Ramirez}, {Young}, {Rieke},
  {Gorlova}, {Muzerolle}, {Slesnick}, \& {Skrutskie}}]{Stauffer2007.1}
{Stauffer}, J.~R., {Hartmann}, L.~W., {Fazio}, G.~G., {et~al.} 2007, \apjs,
  172, 663

\bibitem[{{Stauffer} {et~al.}(2003){Stauffer}, {Jones}, {Backman}, {Hartmann},
  {Barrado y Navascu{\'e}s}, {Pinsonneault}, {Terndrup}, \&
  {Muench}}]{Stauffer2003.1}
{Stauffer}, J.~R., {Jones}, B.~F., {Backman}, D., {et~al.} 2003, \aj, 126, 833

\bibitem[{{Stauffer} {et~al.}(1998{\natexlab{a}}){Stauffer}, {Schild}, {Barrado
  y Navascues}, {Backman}, {Angelova}, {Kirkpatrick}, {Hambly}, \&
  {Vanzi}}]{Stauffer1998.1}
{Stauffer}, J.~R., {Schild}, R., {Barrado y Navascues}, D., {et~al.}
  1998{\natexlab{a}}, \apj, 504, 805

\bibitem[{{Stauffer} {et~al.}(1998{\natexlab{b}}){Stauffer}, {Schultz}, \&
  {Kirkpatrick}}]{Stauffer1998.2}
{Stauffer}, J.~R., {Schultz}, G., \& {Kirkpatrick}, J.~D. 1998{\natexlab{b}},
  \apjl, 499, L199+

\bibitem[{{Stuik} {et~al.}(1997){Stuik}, {Bruls}, \&
  {Rutten}}]{Stuik1997-Pleiades-Li-K-Activityspread}
{Stuik}, R., {Bruls}, J.~H.~M.~J., \& {Rutten}, R.~J. 1997, \aap, 322, 911

\bibitem[{{Terndrup} {et~al.}(2000){Terndrup}, {Stauffer}, {Pinsonneault},
  {Sills}, {Yuan}, {Jones}, {Fischer}, \&
  {Krishnamurthi}}]{Terndrup2000-Rotation-Pleiades-Hyades}
{Terndrup}, D.~M., {Stauffer}, J.~R., {Pinsonneault}, M.~H., {et~al.} 2000,
  \aj, 119, 1303

\bibitem[{{Thorburn} {et~al.}(1993){Thorburn}, {Hobbs}, {Deliyannis}, \&
  {Pinsonneault}}]{Thorburn1993-Li-Hyades}
{Thorburn}, J.~A., {Hobbs}, L.~M., {Deliyannis}, C.~P., \& {Pinsonneault},
  M.~H. 1993, \apj, 415, 150

\bibitem[{{van Leeuwen}(1999)}]{vanLeeuwen1999-Hipparcos-9clusters}
{van Leeuwen}, F. 1999, \aap, 341, L71

\bibitem[{{van Leeuwen}(2009)}]{vanLeeuwen2009.1}
{van Leeuwen}, F. 2009, \aap, 500, 505

\end{thebibliography}

%%%%%%%%%%%%%%%%%%%%%%%%%%%%%%%%%%%%%%%%%%%%%
%%%%%%%%%%%%%%%%%%%%%%%%%%%%%%%%%%%%%%%%%%%%%
%
%  FIGURES
%
%%%%%%%%%%%%%%%%%%%%%%%%%%%%%%%%%%%%%%%%%%%%%
%%%%%%%%%%%%%%%%%%%%%%%%%%%%%%%%%%%%%%%%%%%%%
%%%%%%%%%%%%%%%%%%%%%%%%%%%%%%%%%%%%%%%%%%%%%
%%%%%%%%%%%%%%%%%%%%%%%%%%%%%%%%%%%%%%%%%%%%%
%%%%%%%%%%%%%%%%%%%%%%%%%%%%%%%%%%%%%%%%%%%%%
\clearpage

%%%%%%%%%%%%%%%%%%%%%%%%%%%%%%%%%%%%%%%   FIGURE 
%%%%%%%%%%%%%%%%%%%%%%%%%%%%%%%%%%%%%%%

% \clearpage

%%%%%%%%%%%%%%%%%%%%%%%%%%%%%%%%%%%%%%%   FIGURE 
%%%%%%%%%%%%%%%%%%%%%%%%%%%%%%%%%%%%%%%

%%%%%%%%%%%%%%%%%%%%%%%%%%%%%%%%%%%%%%%   FIGURE 
%%%%%%%%%%%%%%%%%%%%%%%%%%%%%%%%%%%%%%%

%%%%%%%%%%%%%%%%%%%%%%%%%%%%%%%%%%%%%%%   FIGURE 
%%%%%%%%%%%%%%%%%%%%%%%%%%%%%%%%%%%%%%%

%%%%%%%%%%%%%%%%%%%%%%%%%%%%%%%%%%%%%%%   FIGURE 
%%%%%%%%%%%%%%%%%%%%%%%%%%%%%%%%%%%%%%%

 \clearpage

%%%%%%%%%%%%%%%%%%%%%%%%%%%%%%%%%%%%%%%   FIGURE 
%%%%%%%%%%%%%%%%%%%%%%%%%%%%%%%%%%%%%%%

\clearpage

%%%%%%%%%%%%%%%%%%%%%%%%%%%%%%%%%%%%%%%   FIGURE 
%%%%%%%%%%%%%%%%%%%%%%%%%%%%%%%%%%%%%%%

\clearpage

%%%%%%%%%%%%%%%%%%%%%%%%%%%%%%%%%%%%%%%   FIGURE 
%%%%%%%%%%%%%%%%%%%%%%%%%%%%%%%%%%%%%%%

%%%%%%%%%%%%%%%%%%%%%%%%%%%%%%%%%%%%%%%%%%%%%
%%%%%%%%%%%%%%%%%%%%%%%%%%%%%%%%%%%%%%%%%%%%%
%
%  TABLES
%
%%%%%%%%%%%%%%%%%%%%%%%%%%%%%%%%%%%%%%%%%%%%%
%%%%%%%%%%%%%%%%%%%%%%%%%%%%%%%%%%%%%%%%%%%%%
%%%%%%%%%%%%%%%%%%%%%%%%%%%%%%%%%%%%%%%%%%%%%
%%x%%%%%%%%%%%%%%%%%%%%%%%%%%%%%%%%%%%%%%%%%%%
%%%%%%%%%%%%%%%%%%%%%%%%%%%%%%%%%%%%%%%%%%%%%
\clearpage

\setcounter{table}{0}
%
%  TABLE 1: Lithium compilation
%
%
%_____________________________________________________________
%                              Table longer than a single page
%                                             and in landscape 
%  In the preamble, use:       \usepackage{lscape}
%-------------------------------------------------------------
% All long tables will be placed automatically at the end, after
%                                        \end{thebibliography}
%
%
\clearpage
\onecolumn
\begin{longtab}
\tiny
\begin{landscape}
\begin{longtable}{lllrrlrrrrrrrrrrrrl}
\caption{\label{tab:LiCompilation} Complete compilation of the Lithium equivalent width.}\\
\hline
{Name}      &
{RA}        &
{DEC}       &
{WLi-F}$^1$ & 
{eWLi-F}    &
{Source}    &
{WLi}$^2$   & 
{WLi}$^3$   & 
{eWLi}      &
{WLi}$^4$   &
{WLi}$^5$   &
{WLi}$^6$   &
{WLi}$^7$   &
{eWLi}      & 
\,          &
{WLi}$^8$   & 
{eWLi}      & 
{SNR}       &
{SNR}       \\
%%%
\cline{8-9} \cline{13-14}  \cline{16-17}
%%%
            &
(deg)       &
(deg)       &
(m\AA)      &
(m\AA)      &
            &
Ki10        &
%Jo96        &
%Jo96        &
 \multicolumn{2}{c}{Jo96} &
Op97        &
So93        &
Ma07        &
%Je99        &
%Je99        &
 \multicolumn{2}{c}{Je99} &
%GL94        &
%GL94        &
\,           &
\multicolumn{2}{c}{GL94} &
Ki10        &
Ma07        \\
\hline
%
% \hline
\endfirsthead
\caption{continued.}\\
\hline
{NameF}     &
{RAdegF}    &
{DECdegF}   &
{WLi-F}     &
{eWLi-F}    &
{Source}    &
{WLi}$^2$   & 
{WLi}$^3$   & 
{eWLi}      & 
{WLi}$^4$   & 
{WLi}$^5$   & 
{WLi}$^6$   & 
{WLi}$^7$   & 
{eWLi}      &
\,          &
{WLi}$^8$   & 
{eWLi}      &
{SNR}       &
{SNR}       \\
%%%
\cline{8-9} \cline{13-14}  \cline{16-17}
%%%
            &
(deg)       &
(deg)       &
(m\AA)      &
(m\AA)      &
            &
K10         &
%Jo96        &
%Jo96        &
 \multicolumn{2}{c}{Jo96} &
Op97        &
So93        &
Ma07        &
%Je99        &
%Je99        &
 \multicolumn{2}{c}{Je99} &
\,          &
%GL94        &
%GL94        &
 \multicolumn{2}{c}{GL94} &
Ki10        &
Ma07        \\

\hline
\endhead
\hline
\endfoot
\hline
% NameF           RAdegF      DECdegF         WLi-F         eWLiF   S-F    WLiK10     WLiJ96     eJ96    WLiO97     WLiS93      WLiM07             WLiJ99     eWJ99     WLiG94     eG94     S/NK10    S/NM07
 BPL163          & 56.993000  & 22.114194  &  $\le$70    &    --  & Da15 &    --   &      --  &   --  &       --  &     --  &      --           &      --  &   --  &\,&        --  &    --  &     --  &    -- \\
 BPL327          & 58.846125  & 24.818139  & $\le$200    &    --  & Da15 &    --   &      --  &   --  &       --  &     --  &      --           &      --  &   --  &\,&        --  &    --  &     --  &    -- \\
 LZJ-50          & 55.983292  & 25.607083  &  $\le$70    &    --  & Da15 &    --   &      --  &   --  &       --  &     --  &      --           &      --  &   --  &\,&        --  &    --  &     --  &    -- \\
 PPL-14          & 56.142833  & 23.856917  &      350    &   90   & Da15 &    --   &      --  &   --  &       --  &     --  &      --           &      --  &   --  &\,&        --  &    --  &     --  &    -- \\
 AK I-1-317      & 58.589973  & 24.075645  &      78     & 10     & Ma07 &    --   &      --  &   --  &       --  &     --  &    78             &      --  &   --  &\,&        --  &    --  &     --  &   230 \\
 AK II-437       & 54.921543  & 23.290886  &      142    & 10     & Ma07 &    --   &      --  &   --  &       --  &     --  &    142            &      --  &   --  &\,&        --  &    --  &     --  &   145 \\
 AK III-756      & 55.400673  & 25.619331  &      120    & 10     & Ma07 &    --   &      --  &   --  &       --  &     --  &    120            &      --  &   --  &\,&        --  &    --  &     --  &   155 \\
 AK V-595        & 60.934029  & 22.944309  &      32     & 10     & Ma07 &    --   &      --  &   --  &       --  &     --  &    32             &      --  &   --  &\,&        --  &    --  &     --  &   200 \\
\hline                                                                                                                                                                      
\end{longtable}
% $\,$
%
\begin{flushleft}
Lithium references by priority: % (*** TO BE REORDERED!!! ***):
Ki10 = \cite{King2010-Li-K-Scatter-Pleiades},
Jo96 = \cite{Jones1996.1},
Op97 = \cite{Oppenheimer1997.1},
So93 = \cite{Soderblom1993-Li-Pleiades},
Ma07 = \cite{Margheim2007.1}, 
Je99 = \cite{Jeffries1999-Li-Pleiades}, 
GL94 = \cite{GarciaLopez1994.1},
\\
$^1$ Selected lithium equivalent width.\\
$^2$ The errors can be estimated as 10 m\AA{} in \cite{King2010-Li-K-Scatter-Pleiades}.\\
$^3$ The FeI6707.4 \AA{} line does not contributed to the listed W(Li). The errors can be estimated as 5 and 20 m\AA{} for slow and fast rotators (\citealp{Jones1996.1}).\\
$^4$ If present, the LiI6708 \AA{} feature dominates the region for these mid-M dwarfs in  \cite{Oppenheimer1997.1}.\\
$^5$ The FeI6707.4 \AA{} line does not contributed to the listed W(Li). The errors can be estimated as 10 m\AA{} in \cite{Soderblom1993-Li-Pleiades}, based on their comparison with previous data.\\
$^6$ R$\sim$12500. W(Li) includes the contribution by FeI6707.4 \AA. The errors have been estimated as 10 m for the \AA\cite{Margheim2007.1} data. The quoted internal errors are 1-2 m\AA\\
$^7$ SNR=80-110, R$\sim$14500. W(Li) after removing the FeI contribution, as listed in \cite{Jeffries1999-Li-Pleiades}. \\
$^8$ R$\sim$10000. We include the   W(Li) as listed in \cite{GarciaLopez1994.1}, since the FeI6707.4 \AA{} contribution was eliminated in that paper.\\
$^a$ CFHT-PL-09: W(Li)$\le$60 m\AA{ } in \cite{Dahm2015-Pleiades-LDB}.\\
$^b$ Teide02 =  CFHT-PL-13: (Li)=600 m\AA{ } in  \cite{Stauffer1998.1} and W(Li)=520$\pm$130 m\AA{ }  in \cite{Dahm2015-Pleiades-LDB}.\\
$^c$ HCG0131 is listed as HCG 938 in \cite{Margheim2007.1}, which does not exist. After checking the available data we believe the fiber was allocated to HCG 131.\\
\end{flushleft}
\end{landscape}
\end{longtab}

\clearpage

\setcounter{table}{1}
%
%  TABLE 2: Astrometry: Pleiades lithium sample and single
%
 %
%_____________________________________________________________
%                              Table longer than a single page
%                                             and in landscape 
%  In the preamble, use:       \usepackage{lscape}
%-------------------------------------------------------------
% All long tables will be placed automatically at the end, after
%                                        \end{thebibliography}
%
\clearpage
\onecolumn
\begin{longtab}
\begin{landscape}
\begin{longtable}{lrlrrrrrrrl}
\caption{\label{tab:TABastrometry} Proper motions and membership probabilities for the Pleiades lithium sample. Only single stars with membership probability larger than 0.75, after removing binaries and suspected binaries.}\\
\hline
{Name}      &
{RA}        &
{DEC}       &
{pmRA}      &
{pmDEC}     &
{epmRA}     &
{epmDE}     &
{Prob}$^1$     &
{Prob}$^2$      &
{Source}$^3$      &
{Other names}  \\
{}      &
{(deg)}        &
{(deg)}       &
{(arcsec)}      &
{(arcsec)}     &
{(arcsec)}     &
{(arcsec)}     &
{All}     &
{PM}      &
{PM}      &
{}  \\
\hline
\endfirsthead
\caption{continued.}\\
\hline
{Name}      &
{RA}        &
{DEC}       &
{pmRA}      &
{pmDEC}     &
{epmRA}     &
{epmDE}     &
{Prob}$^1$     &
{Prob}$^2$      &
{Source}$^3$      &
{Other names}  \\
{}      &
{(deg)}        &
{(deg)}       &
{(arcsec)}      &
{(arcsec)}     &
{(arcsec)}     &
{(arcsec)}     &
{All}     &
{PM}      &
{PM}      &
{}  \\
\hline
\endhead
\hline
\endfoot
\hline
% Name          RA         DEC         pmRA    pmDEC    epmRA  epmDEC PrAll PrPM    SoPM    OtherName                      
\hline                                                                                                                                                                                                                                                                                                     
BPL163     &   56.991825 & 22.114128 & 17.74 & -39.84 & 0.19 & 0.19 & 1.0  & 0.9  & D, ?: D &                         \\ % 
BPL327     &   58.846134 & 24.818094 & 20.17 & -40.73 & 0.38 & 0.38 & 1.0  & 0.6  & D, ?: D & HHJ I-14, PLZJ78        \\ % 
LZJ-50     &   55.98334  & 25.607027 & 16.97 & -41.44 & 0.32 & 0.32 & 1.0  & 0.93 & D, ?: D &                         \\ % 
PPL-14     &   56.142925 & 23.856792 & 17.0  & -38.82 & 0.45 & 0.45 & 1.0  & 0.9  & D, ?: D &                         \\
AK I-1-317 &   58.589973 & 24.075645 & 19.7  & -46.0  & 0.9  & 1.0  & 1.0  & 0.97 & D, T: T & AK1A317 HD24463                     \\   %   
AK II-437  &   54.921543 & 23.290886 & 22.2  & -43.7  & 1.0  & 1.1  & 1.0  & 0.92 & D, T: T & PELS017 HD22680 AK 2437             \\   %   
AK III-756 &   55.400673 & 25.619331 & 21.1  & -44.7  & 1.7  & 1.7  & 1.0  & 0.95 & D, T: T & PELS023 DH181 AK 3756               \\   %   
CFHT-PL-11 &   56.9125   & 24.606138 & 17.29 & -41.97 & 0.14 & 0.14 & 1.0  & 0.92 & D, ?: D & Roque16 BRB12 BPL152                \\   %   
CFHT-PL-12 &   58.479584 & 23.393723 & 14.72 & -38.74 & 0.14 & 0.14 & 1.0  & 0.89 & D, ?: D & BPL294 PLIZ6 BRB9 PLZJ9             \\   %   
CFHT-PL-16 &   56.146667 & 25.228695 & 15.63 & -42.0  & 0.18 & 0.18 & 1.0  & 0.9  & D, ?: D & PLIZ9                               \\   %   
Calar-03   &   57.856667 & 23.755722 & 15.47 & -40.72 & 0.23 & 0.23 & 1.0  & 0.92 & D, ?: D & CFHT-PL-21 BRB14 PLIZ12 BPL235      \\   %   
PPl-15     &   57.020084 & 23.65889  & 11.75 & -36.47 & 0.65 & 0.65 & 1.0  & 0.36 & D, ?: D & NPL35 SHF48 IPMBD23                 \\   %   
Pels124    &   53.88203  & 22.823627 & 21.0  & -44.4  & 1.2  & 1.3  & 1.0  & 0.96 & D, T: T & PELS124                             \\   %   
Pels192    &   58.870853 & 23.772451 & 14.61 & -40.75 & 0.53 & 0.53 & 0.99 & 0.88 & D, ?: D & PELS192 HCG498                      \\   %   
Roque13    &   56.460835 & 24.150833 & 15.2  & -38.17 & 0.46 & 0.46 & 1.0  & 0.89 & D, ?: D & BPL79 SHF19                         \\   %   
Teide01    &   56.82458  & 24.375528 & 14.39 & -38.91 & 0.65 & 0.65 & 1.0  & 0.87 & D, ?: D & BPL137 NPL39                        \\   %   
Teide02    &   58.027916 & 24.266945 & 18.62 & -42.4  & 0.14 & 0.14 & 1.0  & 0.85 & D, ?: D & CFHT-PL-13 BPL254 PLIZ3 PLZJ46 BRB11\\   %   
HCG0332    &   57.110798 & 23.191612 & 23.99 & -36.63 & 0.14 & 0.14 & 0.89 & 0.07 & D, ?: D & HCG332 SK362 HHJ339                 \\   %   
HII0025    &   55.729626 & 24.493065 & 19.9  & -43.2  & 1.1  & 1.1  & 1.0  & 0.96 & D, T: T & HII25 HD23061 TrR11                 \\   %   
HII0034    &   55.76223  & 24.669737 & 21.8  & -48.8  & 3.7  & 3.7  & 0.93 & 0.88 & D, T: T & HII34 DH250                         \\   %   
HII0097    &   55.860924 & 24.994333 & 20.97 & -39.43 & 0.25 & 0.25 & 1.0  & 0.33 & D, ?: D & HII97 HCG108 DH266                  \\   %   
HII0120    &   55.88314  & 23.674074 & 20.7  & -46.7  & 1.5  & 1.5  & 1.0  & 0.95 & D, T: T & HII120 Tr37                         \\   %   
HII0129    &   55.89337  & 23.761917 & 20.2  & -45.1  & 2.4  & 2.5  & 0.99 & 0.94 & D, T: T & HII129 Tr041 DH271                  \\   %   
HII0133    &   55.903843 & 24.393944 & 15.81 & -38.9  & 0.15 & 0.15 & 0.98 & 0.92 & D, ?: D & HII133 HCG118 SK625 DH277           \\   %   
HII0152    &   55.907204 & 23.53601  & 19.4  & -47.8  & 2.0  & 2.1  & 0.99 & 0.94 & D, T: T & HII152 Tr46                         \\   %   
HII0158    &   55.93021  & 24.37455  & 18.0  & -43.9  & 1.0  & 0.9  & 1.0  & 0.94 & ?, T: T & HII158 HD23156 Tr51                 \\   %   
HII0164    &   55.928608 & 23.5948   & 21.4  & -45.6  & 1.1  & 1.0  & 0.99 & 0.96 & D, T: T & HII164 HD23158 Tr53                 \\   %   
HII0174    &   55.951378 & 25.004387 & 22.0  & -45.7  & 2.0  & 2.1  & 1.0  & 0.93 & D, T: T & HII174 Tr56                         \\   %   
HII0193    &   55.961273 & 24.247465 & 22.0  & -47.4  & 1.8  & 1.9  & 0.99 & 0.92 & D, T: T & HII193 Tr060                        \\   %   
HII0250    &   56.01768  & 24.989822 & 16.6  & -47.1  & 1.8  & 1.8  & 0.99 & 0.88 & D, T: T & HII250 Tr080                        \\   %   
HII0253    &   56.014748 & 24.504206 & 24.4  & -45.0  & 1.9  & 2.1  & 0.99 & 0.81 & D, T: T & HII253 Tr081 DH309                  \\   %   
HII0263    &   56.02017  & 24.275501 & 14.59 & -47.06 & 0.42 & 0.42 & 1.0  & 0.31 & D, ?: D & HII263 Tr089                        \\   %   
HII0293    &   56.05798  & 24.779388 & 21.2  & -46.0  & 1.8  & 1.8  & 1.0  & 0.95 & D, T: T & HII293 Tr098                        \\   %   
HII0324    &   56.09127  & 24.768408 & 15.72 & -43.32 & 0.68 & 0.68 & 1.0  & 0.83 & D, ?: D & HII324 HCG147 DH328                 \\   %   
HII0344    &   56.107113 & 24.394674 & 22.7  & -48.7  & 1.1  & 1.0  & 0.99 & 0.85 & D, T: T & HII344 HD23246 Tr121                \\   %   
HII0345    &   56.109486 & 24.589705 & 8.79  & -56.42 & 0.69 & 0.69 & 1.0  & 0.0  & D, ?: D & HII345 Tr122                        \\   %   
HII0380    &   56.1559   & 25.137793 & 21.0  & -38.87 & 2.53 & 2.53 & 1.0  & 0.6  & D, ?: D & HII380 DH349                        \\   %   
HII0405    &   56.1698   & 24.818542 & 17.0  & -46.3  & 1.5  & 1.5  & 1.0  & 0.91 & ?, T: T & HII405 HD23269 Tr131                \\   %   
HII0430    &   56.18325  & 24.231215 & 22.12 & -43.55 & 0.75 & 0.75 & 1.0  & 0.24 & D, ?: D & HII430 Tr137                        \\   %   
HII0470    &   56.21348  & 23.268969 & 20.4  & -42.2  & 0.9  & 0.9  & 0.99 & 0.92 & D, T: T & HII470 HD23289 Tr149                \\   %   
HII0489    &   56.23499  & 24.432623 & 14.36 & -44.23 & 0.63 & 0.63 & 1.0  & 0.57 & D, ?: D & HII489 Tr153                        \\   %   
HII0514    &   56.26669  & 25.257841 & 18.0  & -43.5  & 1.6  & 1.6  & 0.98 & 0.93 & D, T: T & HII514 DH370                        \\   %   
HII0530    &   56.27201  & 23.702707 & 20.3  & -44.5  & 1.0  & 1.0  & 1.0  & 0.97 & D, T: T & HII530 HD23326 Tr163                \\   %   
HII0531    &   56.27723  & 24.263523 & 20.0  & -45.0  & 1.1  & 1.0  & 1.0  & 0.97 & ?, T: T & HII531 HD23325 Tr162                \\   %   
HII0627    &   56.350533 & 24.885977 & 19.9  & -43.8  & 1.2  & 1.2  & 1.0  & 0.96 & ?, T: T & HII627 HD23352 Tr187                \\   %   
HII0636    &   56.34248  & 23.471731 & 16.26 & -39.34 & 0.23 & 0.23 & 1.0  & 0.93 & D, ?: D & HII636 Tr190 DH390                  \\   %   
HII0652    &   56.358902 & 24.03514  & 18.1  & -45.1  & 1.0  & 1.0  & 1.0  & 0.95 & D, T: T & HII652 HD23361 Tr195                \\   %   
HII0676    &   56.373203 & 23.760529 & 15.2  & -41.75 & 0.16 & 0.16 & 1.0  & 0.89 & D, ?: D & HII676 Tr203 HCG190 DH395           \\   %   
HII0686    &   56.387222 & 24.303242 & 15.33 & -38.89 & 0.43 & 0.43 & 1.0  & 0.91 & D, ?: D & HII686 Tr206 HCG192                 \\   %   
HII0708    &   56.397484 & 24.08321  & 21.81 & -37.44 & 0.8  & 0.8  & 1.0  & 0.2  & D, ?: D & HII708 Tr213                        \\   %   
HII0740    &   56.426872 & 25.05709  & 10.21 & -34.36 & 0.88 & 0.87 & 0.95 & 0.12 & D, ?: D & HII740                              \\   %   
HII0746    &   56.424374 & 24.431513 & 19.73 & -44.07 & 0.75 & 0.75 & 1.0  & 0.64 & D, ?: D & HII746                              \\   %   
HII0879    &   56.52706  & 24.56742  & 16.45 & -43.22 & 0.48 & 0.48 & 1.0  & 0.86 & D, ?: D & HII879 Tr253 DH420                  \\   %   
HII0882    &   56.517178 & 23.405537 & 22.23 & -37.87 & 1.04 & 1.03 & 0.96 & 0.19 & D, ?: D & HII882 Tr256 DH418                  \\   %   
HII0883    &   56.528744 & 24.562807 & 16.73 & -42.06 & 0.6  & 0.6  & 1.0  & 0.91 & D, ?: D & HII883 Tr257 HCG208                 \\   %   
HII0916    &   56.54894  & 24.622314 & 14.79 & -51.74 & 0.46 & 0.46 & 1.0  & 0.09 & D, ?: D & HII916 Tr268 DH428                  \\   %   
HII0923    &   56.541874 & 23.340015 & 22.0  & -51.8  & 0.9  & 0.9  & 0.9  & 0.72 & D, T: T & HII923 Tr270 DH427                  \\   %   
HII0974    &   56.58531  & 24.78548  & 13.42 & -42.07 & 1.18 & 1.19 & 1.0  & 0.73 & D, ?: D & HII974 Tr284 SK475 DH438            \\   %   
HII0996    &   56.59448  & 24.570177 & 10.21 & -53.29 & 1.89 & 1.89 & 1.0  & 0.03 & D, ?: D & HII996 HD282963 Tr289               \\   %   
HII1005    &   56.60015  & 24.361542 & 15.5  & -38.77 & 0.46 & 0.46 & 1.0  & 0.91 & D, ?: D &                                     \\   %   
HII1015    &   56.613968 & 25.135563 & 20.0  & -44.6  & 1.4  & 1.4  & 1.0  & 0.97 & ?, T: T & HII1015 HD282952 TrR13 DH448        \\   %   
HII1032    &   56.618378 & 24.433918 & 21.9  & -40.29 & 1.23 & 1.23 & 1.0  & 0.34 & D, ?: D & HII1032 Tr296 DH450                 \\   %   
HII1095    &   56.65739  & 24.747698 & 13.32 & -57.38 & 1.52 & 1.54 & 1.0  & 0.01 & D, ?: D & HII1095 Tr316 DH462                 \\   %   
HII1110    &   56.662018 & 24.520338 & 16.94 & -41.49 & 0.4  & 0.4  & 1.0  & 0.93 & D, ?: D & HII1110 Tr322                       \\   %   
HII1122    &   56.663857 & 24.103243 & 13.54 & -51.28 & 1.34 & 1.33 & 1.0  & 0.11 & D, ?: D & HII1122 HD23511 Tr327               \\   %   
HII1124    &   56.664104 & 24.029688 & 16.57 & -40.21 & 0.26 & 0.26 & 1.0  & 0.94 & D, ?: D & HII1124 Tr329 DH463                 \\   %   
HII1132    &   56.659985 & 22.919804 & 21.1  & -43.0  & 1.1  & 1.0  & 1.0  & 0.94 & D, T: T & HII1132 HD23514                     \\   %   
HII1139    &   56.666656 & 23.11037  & 17.2  & -46.5  & 1.2  & 1.2  & 1.0  & 0.91 & D, T: T & HII1139 HD23513 Tr331               \\   %   
HII1182    &   56.696087 & 22.914593 & 19.8  & -46.6  & 1.4  & 1.4  & 1.0  & 0.96 & D, T: T & HII1182                             \\   %   
HII1200    &   56.71057  & 23.239195 & 17.6  & -37.9  & 1.0  & 1.0  & 0.94 & 0.22 & D, T: T & HII1200 Tr346                       \\   %   
HII1207    &   56.72883  & 24.796349 & 17.7  & -45.8  & 1.8  & 1.8  & 0.98 & 0.93 & D, T: T & HII1207 HD282962 Tr348 DH477        \\   %   
HII1215    &   56.72391  & 23.5836   & 20.2  & -43.9  & 1.6  & 1.7  & 0.99 & 0.95 & D, T: T & HII1215 Tr350                       \\   %   
HII1220    &   56.72192  & 22.880943 & 16.47 & -43.54 & 0.5  & 0.5  & 1.0  & 0.84 & D, ?: D & HII1220                             \\   %   
HII1284    &   56.76755  & 23.995174 & 19.3  & -44.4  & 1.1  & 1.0  & 1.0  & 0.97 & D, T: T & HII1284 HD23585 Tr365               \\   %   
HII1298    &   56.77825  & 23.71517  & 10.02 & -41.36 & 0.29 & 0.29 & 1.0  & 0.25 & D, ?: D & HII1298 Tr369                       \\   %   
HII1305    &   56.78061  & 23.22634  & 16.64 & -41.2  & 0.18 & 0.18 & 1.0  & 0.93 & D, ?: D & HII1305 HCG255 DH493                \\   %   
HII1309    &   56.791893 & 24.276672 & 20.4  & -44.6  & 1.1  & 1.1  & 1.0  & 0.97 & D, T: T & HII1309 HD23584 Tr372               \\   %   
HII1332    &   56.806374 & 23.71431  & 14.35 & -40.36 & 0.32 & 0.32 & 1.0  & 0.87 & D, ?: D & HII1332 Tr378                       \\   %   
HII1362    &   56.830624 & 24.139126 & 19.2  & -43.0  & 1.2  & 1.1  & 1.0  & 0.94 & D, T: T & HII1362 HD23607 Tr390               \\   %   
HII1407    &   56.845417 & 22.922121 & 20.9  & -47.9  & 1.1  & 1.0  & 1.0  & 0.93 & ?, T: T & HII1407 HD23610 TrS108              \\   %   
HII1454    &   56.89032  & 24.684223 & 14.87 & -42.39 & 0.54 & 0.54 & 1.0  & 0.83 & D, ?: D & HII1454 HCG278 DH521                \\   %   
HII1514    &   56.918537 & 24.364594 & 12.97 & -40.48 & 0.62 & 0.62 & 1.0  & 0.69 & D, ?: D & HII1514 HD282967 Tr430              \\   %   
HII1531    &   56.922657 & 23.971945 & 17.64 & -38.44 & 0.35 & 0.35 & 1.0  & 0.85 & D, ?: D & HII1531 Tr435 HCG285 DH532          \\   %   
HII1532    &   56.92158  & 23.740294 & 17.03 & -39.34 & 0.33 & 0.33 & 0.99 & 0.92 & D, ?: D & HII1532 HCG286 SK405 B270           \\   %   
HII1593    &   56.950462 & 23.21814  & 19.0  & -45.2  & 2.3  & 2.4  & 0.95 & 0.94 & D, T: T & HII1593 Tr444                       \\   %   
HII1613    &   56.968838 & 23.941282 & 16.15 & -51.94 & 0.94 & 0.94 & 1.0  & 0.09 & D, ?: D & HII1613 HD282973 Tr448              \\   %   
HII1756    &   57.045776 & 23.507023 & 17.87 & -38.56 & 0.19 & 0.19 & 0.99 & 0.84 & D, ?: D & HII1756 SK378 DH557                 \\   %   
HII1776    &   57.07372  & 25.047886 & 10.68 & -56.07 & 1.32 & 1.32 & 1.0  & 0.01 & D, ?: D & HII1776 Tr495 DH567                 \\   %   
HII1794    &   57.071342 & 23.890385 & 11.52 & -44.12 & 1.26 & 1.26 & 1.0  & 0.34 & D, ?: D & HII1794 HD282972 Tr499              \\   %   
HII1797    &   57.070457 & 23.636806 & 21.06 & -30.59 & 1.65 & 1.65 & 1.0  & 0.05 & D, ?: D & HII1797 Tr500                       \\   %   
HII1856    &   57.109024 & 24.048445 & 21.32 & -44.71 & 0.83 & 0.82 & 1.0  & 0.32 & D, ?: D & HII1856 HD282971 Tr514              \\   %   
HII1883    &   57.116787 & 23.300774 & 21.45 & -41.83 & 0.31 & 0.31 & 1.0  & 0.33 & D, ?: D & HII1883 Tr520 HCG331 DH579          \\   %   
HII1924    &   57.143806 & 23.434818 & 14.06 & -39.77 & 1.43 & 1.43 & 1.0  & 0.84 & D, ?: D & HII1924                             \\   %   
HII2016    &   57.189312 & 23.338873 & 10.47 & -42.09 & 1.13 & 1.13 & 0.97 & 0.3  & D, ?: D & HII2016 Tr557 HCG344                \\   %   
HII2034    &   57.20551  & 23.97732  & 15.98 & -37.61 & 0.4  & 0.4  & 1.0  & 0.86 & D, ?: D & HII2034 Tr563 HCG348                \\   %   
HII2106    &   57.24369  & 23.20123  & 15.9  & -38.91 & 0.72 & 0.71 & 1.0  & 0.91 & D, ?: D & HII2106                             \\   %   
HII2126    &   57.259705 & 23.252468 & 12.77 & -39.06 & 1.43 & 1.41 & 1.0  & 0.72 & D, ?: D & HII2126 Tr584                       \\   %   
HII2244    &   57.335793 & 24.77667  & 13.38 & -55.0  & 5.6  & 5.6  & 0.99 & 0.2  & D, ?: D & HII2244 Tr616 HCG369 DH613          \\   %   
HII2284    &   57.350216 & 23.839287 & 14.24 & -44.01 & 1.16 & 1.16 & 1.0  & 0.65 & D, ?: D & HII2284 Tr628                       \\   %   
HII2311    &   57.369747 & 23.712238 & 15.89 & -41.04 & 0.44 & 0.44 & 1.0  & 0.93 & D, ?: D & HII2311                             \\   %   
HII2341    &   57.388004 & 23.795427 & 18.18 & -36.44 & 1.32 & 1.32 & 1.0  & 0.61 & D, ?: D & HII2341 Tr649                       \\   %   
HII2345    &   57.38634  & 23.380413 & 19.7  & -43.6  & 0.9  & 0.8  & 1.0  & 0.97 & D, T: T & HII2345 HD23912 Tr651               \\   %   
HII2366    &   57.402245 & 24.296135 & 13.75 & -40.82 & 0.89 & 0.89 & 1.0  & 0.81 & D, ?: D & HII2366 Tr657                       \\   %   
HII2415    &   57.420506 & 23.341612 & 19.6  & -43.8  & 0.9  & 0.9  & 1.0  & 0.97 & D, T: T & HII2415 HD23924 Tr670               \\   %   
HII2462    &   57.459827 & 23.705631 & 14.86 & -40.87 & 0.44 & 0.44 & 1.0  & 0.9  & D, ?: D & HII2462 Tr680                       \\   %   
HII2506    &   57.485363 & 23.218641 & 18.5  & -43.8  & 1.1  & 1.1  & 0.99 & 0.95 & D, T: T & HII2506                             \\   %   
HII2588    &   57.551674 & 24.532957 & 15.81 & -42.09 & 0.4  & 0.4  & 1.0  & 0.9  & D, ?: D & HII2588 HCG387 Tr712 DH658          \\   %   
HII2644    &   57.58708  & 24.466755 & 16.39 & -42.25 & 1.15 & 1.15 & 1.0  & 0.89 & D, ?: D & HII2644 Tr727                       \\   %   
HII2665    &   57.588753 & 23.096392 & 20.6  & -44.1  & 1.7  & 1.9  & 0.99 & 0.95 & D, T: T & HII2665                             \\   %   
HII2741    &   57.64405  & 24.507822 & 18.22 & -39.72 & 0.19 & 0.19 & 1.0  & 0.87 & D, ?: D & HII2741 DH675                       \\   %   
HII2786    &   57.666985 & 23.933065 & 16.05 & -45.84 & 0.75 & 0.76 & 1.0  & 0.51 & D, ?: D & HII2786 HD283067 Tr772              \\   %   
HII2870    &   57.714333 & 23.3291   & 14.87 & -42.37 & 0.2  & 0.2  & 1.0  & 0.83 & D, ?: D & HII2870 DH685                       \\   %   
HII2880    &   57.729527 & 24.197466 & 9.71  & -38.92 & 1.02 & 1.02 & 1.0  & 0.22 & D, ?: D & HII2880 Tr795                       \\   %   
HII2927    &   57.77326  & 24.736776 & 14.56 & -40.14 & 0.3  & 0.3  & 1.0  & 0.89 & D, ?: D & HII2927 HCG418 DH693                \\   %   
HII3019    &   57.851673 & 24.087425 & 16.79 & -41.54 & 0.15 & 0.15 & 1.0  & 0.93 & D, ?: D & HII3019 HCG429 DH710                \\   %   
HII3031    &   57.86342  & 24.518692 & 22.4  & -44.7  & 1.2  & 1.1  & 0.99 & 0.93 & D, T: T & HII3031 HD24132 Tr848               \\   %   
HII3063    &   57.874737 & 23.899223 & 22.52 & -44.44 & 0.16 & 0.16 & 0.92 & 0.16 & D, ?: D & HII3063 Tr862 HCG431 DH713          \\   %   
HII3096    &   57.91365  & 24.54894  & 22.75 & -42.07 & 0.47 & 0.47 & 1.0  & 0.18 & D, ?: D & HII3096 Tr879 DH721                 \\   %   
HII3163    &   57.972446 & 24.387018 & 15.15 & -29.05 & 1.23 & 1.23 & 0.86 & 0.04 & D, ?: D & HII3163 Tr911 HCG444 DH725          \\   %   
HII3179    &   57.98689  & 23.901966 & 20.3  & -47.1  & 1.2  & 1.3  & 1.0  & 0.96 & D, T: T & HII3179 HD24194 TrR21               \\   %   
HII3187    &   57.988895 & 23.33944  & 17.89 & -41.16 & 0.23 & 0.23 & 0.98 & 0.91 & D, ?: D & HII3187 DH730                       \\   %   
mhodb3     &   55.475456 & 23.084862 & 17.73 & -41.19 & 0.21 & 0.21 & 1.0  & 0.92 & D, ?: D & DH174                               \\   %   
\hline
\end{longtable}
\begin{flushleft}
$\,$
$^1$ Membership probability based on proper motion and photometry.\\
$^2$ Membership probability based on proper motion only (\citealp{Sarro2014.1}).\\
$^3$ Proper motions from: ``D'' = DANCe (\citealp{Bouy2013-DANCE}), ``T'' = Tycho  (\citealp{Hog2000.1}), and final selection.
\end{flushleft}
\end{landscape} 
\end{longtab}

%\end{document}

\clearpage

\setcounter{table}{2}
%
%  TABLE 3: Lithium, rotation and activity: Pleiades lithium sample and single
%
%
%_____________________________________________________________
%                              Table longer than a single page
%                                             and in landscape 
%  In the preamble, use:       \usepackage{lscape}
%-------------------------------------------------------------
% All long tables will be placed automatically at the end, after
%                                        \end{thebibliography}
%
%
\clearpage
\onecolumn
\begin{longtab}
\tiny
\begin{landscape}
\begin{longtable}{lrrlrllrrllrlrccllrcccll}
\caption{\label{tab:otherdata} Several stellar properties, including lithium, rotation and activity data for the Pleades lithium sample. Only single stars with membership probability larger than 0.75, after removing binaries and suspected binaries.}\\
\hline
{Name}      &
{WLi}$^1$   &
{eWLi}      &
{Ref}       &
{WLi}$^2$   &
{eWLi}      &
{Ref}       &
\multicolumn{2}{c}{Temp}      &
%{Temp}      &
{L$_{bol}$}  &
{eL$_{bol}$} &
{ALi}       &
{eALi}      &
{eALi}      &
{Prot}      &
{eProt}     &
{Amp}       &
{Vsini}     &
{Ref}       &
{Rot}       &
{Li}        &
{Log Lx}    &
{eLx}       &
{Ref}     \\ 
% %%%%%%%%%%%%%%%%%%%%
\cline{2-3}
\cline{5-6}
\cline{8-9}
\cline{15-16}
\cline{22-23}
% %%%%%%%%%%%%%%%%%%%%
   &
(m\AA)   &
(m\AA)   &
         &
(m\AA)   &
(m\AA)   &
         &
(K)$^3$   &
(K)$^4$   &
($L_\odot$)   &
($L_\odot$)   &
   &
(1$\sigma$)   &
(3$\sigma$)   &
(d)   &
(d)   &
(mag)   &
(km/s)   &
   &
 type&
 type&
(erg/s)   &
(erg/s)   &
Xray\\ 
\hline
\endfirsthead
\caption{continued.}\\
\hline\hline
{Name}      &
{WLi}$^1$   &
{eWLi}      &
{Ref}       &
{WLi}$^2$   &
{eWLi}      &
{Ref}       &
\multicolumn{2}{c}{Temp}      &
%{Temp}      &
{L$_{bol}$}  &
{eL$_{bol}$} &
{ALi}       &
{eALi}      &
{eALi}      &
{Prot}      &
{eProt}     &
{Amp}       &
{Vsini}     &
{Ref}       &
{Rot}       &
{Li}        &
{Log Lx}    &
{eLx}       &
{Ref}     \\ 
% %%%%%%%%%%%%%%%%%%%%
\cline{2-3}
\cline{5-6}
\cline{8-9}
\cline{15-16}
\cline{22-23}
% %%%%%%%%%%%%%%%%%%%%
   &
(m\AA)   &
(m\AA)   &
         &
(m\AA)   &
(m\AA)   &
         &
(K)$^3$   &
(K)$^4$   &
($L_\odot$)   &
($L_\odot$)   &
   &
(1$\sigma$)   &
(3$\sigma$)   &
(d)   &
(d)   &
(mag)   &
(km/s)   &
   &
type&
type&
(erg/s)   &
(erg/s)   &
Xray\\ 
\hline
\endhead
\hline
\endfoot
\hline
% Name            WLi     eWLi       SoLi    WLi_F   eWLi  SourceLi   Temp   TeffChi LumBol    erLum      ALi       e1s        e3s       Prot         eProt         Amp      Vsini     Ref    Pclass LiClass LogLx   eLogLx  RefLx %   
%\hline                                                                                                                                                                                          %                                    
AK I-1-317     &    78   &  10     & Ma07  &  78     &  10 &  Ma07  &  6250 & 6305 & 1.95    & 1.8E-1 &   2.973  &  0.029  &  0.085     & --         & --         & --     &   60.0   & Ma07  &-- &     & --     & --    & --   \\ %  
AK II-437      &   142   &  10     & Ma07  &  142    &  10 &  Ma07  &  6250 & 6304 & 1.60    & 1.4E-1 &   3.406  &  0.025  &  0.074     & --         & --         & --     &   29.3   & Qu98  &-- &     & --     & --    & --   \\ %  
AK III-756     &   120   &  10     & Ma07  &  120    &  10 &  Ma07  &  6000 & 5987 & 1.50    & 1.3E-1 &   2.993  &  0.026  &  0.077     & 1.63715883 & 2.9E-5     & 0.0079 &   36.2   & Qu98  & A &     & --     & --    & --   \\ %  
BPL163         & $\le$70 &  --     & Da15  & $\le$70 &  -- &  Da15  &  2800 & 2755 & 0.00169 & 1.9E-4 &$\le$0.353&  --     &  --        & --         & --         & --     & --       & --    &-- &     & --     & --    & --   \\ %  
BPL327         &$\le$200 &  --     & Da15  &$\le$200 &  -- &  Da15  &  2700 & 2681 & 0.00128 & 1.4E-4 &$\le$1.166&  --     &  --        & --         & --         & --     & --       & --    &-- &     & --     & --    & --   \\ %  
CFHT-PL-11     &   500   &  --     & St98  &  500    &  -- &  St98  &  2700 & 2711 & 0.0013  & 1.3E-4 &   2.769  &  0.015  &  0.032     & --         & --         & --     & --       & --    &-- &     & --     & --    & --   \\
\hline                                                                                                                                                                      
\end{longtable}
% $\,$
%
\begin{flushleft}
$\,$
Lithium and rotation sources:
Ba96 = \cite{Basri1996.1},
Da15 = \cite{Dahm2015-Pleiades-LDB},
GL94 = \cite{GarciaLopez1994.1},
Jo96 = \cite{Jones1996.1},
Ki10 = \cite{King2010-Li-K-Scatter-Pleiades},
Ma07 = \cite{Margheim2007.1}, 
Mt98 = \cite{Martin1998.1},
Mt00 = \cite{Martin2000.1},
Op97 = \cite{Oppenheimer1997.1},
Qu98 = \cite{Queloz1998_PleiadesVrot},
Re96 = \cite{Rebolo1996.1}, 
So93 = \cite{Soderblom1993-Li-Pleiades}, So93b = \cite{Soderblom1993-RotationActivity-FGK-Pleiades},
St98 = \cite{Stauffer1998.1},
Te00 = \cite{Terndrup2000-Rotation-Pleiades-Hyades},
St87 = \cite{Stauffer1987-rotation-Pleiades}.
\\
$^1$ W(Li) selection based primarily on the S/N.\\
$^2$ W(Li) selection based primarily on the spectral resolution.\\
$^3$ Effective temperature for the minimum $\chi$$^2$ as derived with VOSA.\\
$^4$ Effective temperature after fitting Teff and $\chi$$^2$ and minimizing this last value.\\
$^{Bin?}$ Suspected binary based on the spectra or the cross-correlation in   \cite{Margheim2007.1}.
\end{flushleft}
\end{landscape}
\end{longtab}

%\cite{Marcy1994.1}, \cite{Jeffries1999-Li-Pleiades}, \cite{Pinfield2003.1},

\clearpage

\end{document}